\documentclass[12pt]{article}\usepackage[hyperfootnotes=false]{hyperref}
\usepackage{epsfig}
\usepackage{float}
\usepackage{empheq}
\usepackage{bbold}
\usepackage[utf8]{inputenc}
\usepackage{amsmath}
\usepackage{caption}
\usepackage{amsmath}
\usepackage{amssymb}
\usepackage{graphicx}
\usepackage{soul}

\newlength{\abstractwidth}
\renewcommand{\thefootnote}{\fnsymbol{footnote}}
\renewcommand{\thanks}[1]{\footnote{#1}} 
\newcommand{\starttext}{
\setcounter{footnote}{0}
\renewcommand{\thefootnote}{\arabic{footnote}}}

\newcommand{\be}{\begin{equation}}
\newcommand{\bea}{\begin{eqnarray}}
\newcommand{\eea}{\end{eqnarray}}
\newcommand{\beq}{\begin{equation}}
\newcommand{\ee}{\end{equation}}

	\newcommand*\widefbox[1]{\fbox{\hspace{2em}#1\hspace{2em}}}

	\def\dsp.{de Sitter space.}

	\def\la{\langle}
	\def\ra{\rangle}
	\def\simleq{\; \raise0.3ex\hbox{$<$\kern-0.75em
			\raise-1.1ex\hbox{$\sim$}}\; }
	\def\simgeq{\; \raise0.3ex\hbox{$>$\kern-0.75em
			\raise-1.1ex\hbox{$\sim$}}\; }
	
	\def\bi{\begin{itemize}}
		\def\ei{\end{itemize}}

	\def\CJ{{\cal{J}}}

	\def\CO{{\cal{O}}}

	\def\bx{{\bar{\chi}}}

	\def\bsub{ \begin{subequations}
			\begin{empheq}[box=\widefbox]{align}  }
			\def\esub{ \end{empheq}
	\end{subequations}}
	
	\def\1{\(  \mathbb{1} \)}

	\def\dk{${\rm DSSYK_{\infty}}$}
	
  \def\bx{\begin{equation} \boxed  }
	
	\makeatletter
	\g@addto@macro\normalsize{%
		\setlength\abovedisplayskip{10pt}
		\setlength\belowdisplayskip{20pt}
		\setlength\abovedisplayshortskip{10pt}
		\setlength\belowdisplayshortskip{20pt}
	}
	\makeatother
	
	\usepackage{color}

	\usepackage{authblk}
	\title{Comments on a Paper by Narovlansky and Verlinde}
	\author[1]{Adel Rahman}
	\author[1,2]{Leonard Susskind}
	\affil[1]{Stanford Institute for Theoretical Physics and Department of Physics\\ Stanford University, Stanford, CA 94305-4060, USA \vspace{1em}}
	\affil[2]{Google, Mountain View, CA}
	\date{}
	\newcommand{\inp} [1] {\left( #1 \right)}
	
	\newcommand{\insb} [1] {\left[ #1 \right]}
	\def\RS{\cite{Susskind:2021esx,Susskind:2022dfz,Susskind:2022bia,Rahman:2022jsf} }
	\usepackage{upgreek}

\begin{document}

		\begin{titlepage}
			\maketitle
			
			\begin{abstract}

The double-scaled infinite temperature limit of the SYK model has been conjectured by Rahman and Susskind (RS)  \cite{Susskind:2021esx,Susskind:2022dfz,Susskind:2022bia,Rahman:2022jsf}, and independently by Verlinde \cite{Verlinde} to be dual to  a certain low dimensional de Sitter space. 
In a recent discussion of this conjecture Narovlansky and Verlinde (NV) \cite{Narovlansky:2023lfz} came to conclusions which radically differ from those of RS.
In particular 
these conclusions disagree by factors which diverge as $N \to \infty$. Among these 
is a mismatch between the scaling of boundary entropy and bulk horizon area. In this note, we point out differences in two key assumptions made by RS and NV which lead to these mismatches, and explain why we think the RS assumptions are correct. When the NV assumptions, which we believe are unwarranted, are replaced by those of RS, the conclusions  match both RS and the standard relation between entropy and area.  

In the process of discussing these, we will shed some light on: the various notions of temperature that appear in the duality; the relationship between Hamiltonian energy and bulk mass; and the location of bulk conical defect states in the spectrum of DSSYK$_{\infty}$.

			\end{abstract}

		\end{titlepage}
		
		\starttext 
		\setcounter{footnote}{0}
		
		\tableofcontents
		
		\section{Introduction}

	\quad 
	It has been conjectured  \cite{Susskind:2021esx,Susskind:2022dfz,Susskind:2022bia,Rahman:2022jsf} by the present authors (RS) and by H. Verlinde \cite{Verlinde} that the double-scaled SYK model at infinite temperature is dual to the $s$-wave sector of ($2+1$)-dimensional de Sitter space  (or, equivalently, to JT gravity with positive cosmological constant \cite{Sybesma:2020fxg,Svesko:2022txo,Rahman:2022jsf}).
Recently,  Narovlansky and Verlinde (NV) \cite{Narovlansky:2023lfz} have put forward a version of the conjecture which at first sight seems to make the same claim as RS. However, RS and  NV 
%
%
%
%
	come to radically different conclusions which disagree by factors which diverge as $N\to \infty.$  
	 Among these is a mismatch, claimed by NV, between the scaling of bulk horizon area and boundary entropy. Specifically, NV claim that
		\begin{equation}
			\frac{\ell_{\mathrm{dS}}}{G} \ \underset{\text{claimed by NV}}{\sim} \ \frac{1}{\lambda}
			\label{NV0}
		\end{equation}
	where $\lambda $ is a finite number in the double-scaled limit. By contrast \RS (and, more generally, the standard holographic relation between bulk area and boundary entropy) would require
		\begin{equation}
			\boxed{\frac{\ell_{\mathrm{dS}}}{G} \ \sim \ N}
			\label{holo}
		\end{equation}
		In both cases the left hand side represents the Gibbons-Hawking entropy (i.e. the horizon ``area" in Planck units)
		\begin{equation}
			\boxed{S_{\mathrm{dS}} \ \equiv \ \frac{2\pi\ell_{\mathrm{dS}}}{4G}}
			\label{Sbulk}
		\end{equation}
		and in \eqref{holo} we have used that the entropy of the double-scaled SYK model goes like $N$ at infinite Boltzmann temperature.
		
It is our view that the standard connection  between horizon area and entropy is the essential foundation of quantum gravity and, to put it mildly, should not be easily given up.  In  this paper we will assume this connection and trace the argument that led NV to \eqref{NV0} down to two key assumptions which are in sharp contradiction with those of RS. 

	The basic methodology employed by NV in the first part of their paper \cite{Narovlansky:2023lfz} is sound and insightful, but we believe 
    that two key errors were made in carrying things out\footnote{These disagreements occur in the first part of \cite{Narovlansky:2023lfz} in Sections 1 and 2. The rest of the paper attempts to holographically define and calculate bulk observables. In this paper we will not address the latter part of \cite{Narovlansky:2023lfz}. The dictionary relating DSSYK$_{\infty}$ and de Sitter correlation functions is an obviously important problem but we have nothing to say about it here (but see \cite{Susskind:2021esx,Rahman:2022jsf,Susskind:2023hnj,Susskind:2023rxm} for comments in this direction).}. When these errors are corrected, the argument of \cite{Narovlansky:2023lfz} becomes consistent with \RS and the standard holographic relation
		\eqref{holo} between bulk area and boundary entropy.

		In the process of unraveling the relation between NV's assumptions and our own we have learned a great deal, in particular new things about:
		\begin{enumerate}
		\item The relationship between the emergent Gibbons-Hawking temperature $T_H$ and the temperature experienced by so-called ``chord operators".
  
            \item The relationship between boundary energy (as defined by the DSSYK Hamiltonian) and bulk mass (as defined by e.g. the $(2 + 1)$-dimensional Schwarzschild-de Sitter metric) at the maximum values for both quantities. We will find that the boundary energy and bulk mass are not generally the same thing. While they agree for small values, their respective maximum values are related by the infinite factor
		\begin{equation}
				\boxed{M_{\mathrm{max}} \ \sim \ p\,E_{\mathrm{max}}}
			\end{equation}

            \item The location of bulk conical deficit states in the spectrum of DSSYK$_{\infty}$: We will find that, in the semiclassical limit (as defined in Appendix \ref{semiclassical} below) bulk states containing conical defects of $O(G^0)$ deficit angle (encoding an $O(1)$ amount of gravitational backreaction) are only found in the non-Gaussian tails of the DSSYK$_{\infty}$ density of states.
            
		\end{enumerate}

\subsection*{Points of Notation and Terminology}

\begin{itemize}

\item
In what follows, we will sometimes quote formulas that we disagree with and believe to be incorrect. To avoid confusion, equations that we believe to be correct (up to irrelevant numerical factors) will be denoted by \fbox{boxing}.

\item  When numerical factors are ignored\footnote{We will for the most part ignore $O(1)$ multiplicative factors since the primary focus will be on dimensional analysis and on multiplicative factors which diverge in the double-scaled/semiclassical limit. The one exception to this rule will be factors which are powers of $\lambda = 2p^2/N$, which will always be left explicit.} we will use the notation 
		\begin{equation}
			A \ \sim \ B
		\end{equation}
		to mean that  the quantity $A$ scales parameterically as the quantity $B$ in some appropriate limit (which limit we mean will be clear from context).

\item  Note that what we will call $\mathcal{J}_0$ (see \eqref{Jst}), NV call  $\mathbb{J}/\sqrt{2}$. From this point on we will use $\CJ_0$ and drop the use of the symbol $\mathbb{J}$. 

\item Following recent works on the double-scaled limit we will denote the $k$-locality parameter of the
		SYK system by $p$ (previously called $q$) and define 
		\begin{equation}
			\boxed{\lambda \equiv \frac{2p^2}{N}}
		\end{equation}
			
\item		We will adapt the terminology introduced in \cite{Susskind:2021omt} in which we will refer to the center of the (e.g. right) static patch as the ``pode"

\item	The notation $\ell_{\mathrm{dS}}$ 
		will denote the de Sitter radius and we work as usual in units in which $\hbar = c  = k_B= 1$.
		
\item The three dimensional Newton constant has units of length and will be denoted $G$. 
		
		\item  The entropy defined by state-counting/infinite temperature Von-Neumann entropy in the SYK theory will be denoted by $S$ and occasionally by $S_{\mathrm{dS}}$ to indicate that it should agree with the maximum entropy \eqref{Sbulk} in de Sitter space.  We will use $S_{\mathrm{GH}}$ (Gibbons-Hawking) to indicate the area of the horizon divided by $4G$,
        \begin{equation}
            S_{\mathrm{GH}} = \frac{\text{horizon area}}{4G}
            \label{SGH}
        \end{equation}
        which will be less than or equal to the maximal value $S_{\mathrm{dS}}$ depending on the choice of state ($S_{\mathrm{GH}} < S_{\mathrm{dS}}$ for states other than the empty static patch/infinite temperature state).

			\end{itemize}	
				
				 In what follows, we will implicitly assume the general framework of static patch horizon holography for de Sitter space discussed in e.g. \cite{Susskind:2021omt}. In other words, we will assume that the hologram lives at the cosmological horizon.

		\section{Why Horizon Area is Entropy in dS}
 		\quad The connection between horizon area  and entropy is a basic tenet of holography. While holography for de Sitter space is somewhat uncharted territory, it is our firm belief that this basic tenet should \emph{not} be thrown out, and should instead be used as a central touchstone. Nevertheless because holography is new to de Sitter space, it is worth giving a small argument for why this should be the case.
		
		A basic requirement for two theories to be dual is that they make the same physical predictions---i.e. determine the same probabilities. In this section we will start with this basic assumption and show that, in de Sitter space, the equivalence of $S_{\mathrm{dS}}$ and the quantum-mechanically defined entropy $S$ is \emph{forced} upon us as a result. 
		
	Dynamics in thermal equilibrium is a theory of fluctuations: on average nothing happens, but every now and then a fluctuation will appear with probability
		\begin{equation}
			\boxed{P \ \sim \ e^{-\Delta S}}
   \label{Pbdy}
		\end{equation}
	where we have specialized to relevant case of infinite temperature and $\Delta S$ denotes the entropy deficit of the state with the fluctuation present relative to the entropy of the infinite temperature equilibrium state (see e.g. \cite{Susskind:2021omt,Chandrasekaran:2022cip}).
		
		Meanwhile, in the bulk we have that 
		\cite{Gibbons:1977mu}
		\begin{equation}
			\boxed{P \ \sim \ e^{-M/T_H}}
		\end{equation} 
		with $M$ the bulk mass of the excitation. But, up to an irrelevant (for our purposes) $O(1)$ coefficient, $M/T_H$ goes like the change in the bulk horizon area  \eqref{Sbulk}:
		\begin{equation}
			\boxed{P \ \sim \ e^{-\Delta S_{\mathrm{GH}}}}
                \label{Pbulk}
		\end{equation}
 		(see Appendix \ref{massApp}---specifically eq. \eqref{SMT}---for details). Here $\Delta S_{\mathrm{GH}}$ represents the change in the Gibbons-Hawking entropy \eqref{SGH}. See also \cite{Chandrasekaran:2022cip}.
		
	Equating the two expressions \eqref{Pbdy}, \eqref{Pbulk} for the probability of a minimum entropy (maximal entropy deficit) state then forces upon us the conclusion that 
		\begin{equation}
			\boxed{S_{\mathrm{dS}} \ \sim \ N}
			\label{central}
			\tag{central assumption}
		\end{equation}
	
		Now that we have explained the central assumption of our analysis---namely the  equivalence of $\text{Area}/4G$ and quantum-mechanical entropy---we will introduce an important concept for our analysis: the separation of scales.
		
    \section{The Separation of Scales }
    \label{SepScales}
    \subsection{The Scales}
	\quad A separation of scales takes place in the semiclassical limit (SCL) of 3D de Sitter space/dS-JT gravity\footnote{By ``dS JT gravity" we mean specifically the system considered by \cite{Sybesma:2020fxg,Svesko:2022txo,Rahman:2022jsf,Narovlansky:2023lfz} with spacelike Dilaton $\Phi = r/\ell_{\mathrm{dS}}$. 
			
		Note that at least two different things are called ``dS-JT" in the literature. Both begin with the same Euclidean dilaton-gravity-matter action 
		\be
		I_{\mathrm{JT}} = \frac{1}{16\pi G}\int_{\mathcal{M}} \mathrm{d}^2x\sqrt{|g|}\,\Phi\inp{R-2/\ell^2}
		- \frac{1}{2}\int_{\mathcal{M}} \mathrm{d}^2x\sqrt{|g|}\,\Phi \,g^{ab}\nabla_a\psi\nabla_b\psi
		\label{action}
		\ee
		(plus possible boundary and topological terms).
		
		The earlier and perhaps more well-known variant of dS-JT gravity was that introduced by \cite{Maldacena:2019cbz,Cotler:2019nbi} who view this theory as describing the near-horizon near-extremal physics of a 4D Schwarzschild de-Sitter black hole. In particular, they take the ``Dilaton" $\Phi$ to fluctuate about a large semiclassical value
		\begin{equation}
			\Phi = \frac{1}{3}S_{\mathrm{dS}} + \phi + \dots \quad\text{with}\quad \frac{\phi}{S_{\mathrm{dS}}} \ll 1
		\end{equation}
		In this scenario matter fields $\psi$ are \emph{uncoupled} from the ``dilaton" $\phi$ and behave as ordinary minimally coupled matter on $(\mathcal{M}, g_{ab})$ at leading order in the semiclassical expansion in $1/S_{\mathrm{dS}}$. 
		
		The variant that we will be concerned with was introduced by \cite{Sybesma:2020fxg,Svesko:2022txo} and revisited recently in \cite{Rahman:2022jsf,Narovlansky:2023lfz} who view this theory as describing the dimensional reduction of 3D de Sitter space in the semiclassical limit. Here the ``Dilaton" $\Phi$ is proprotional to the radial coordinate of the static patch and is $O(1)$; as a result matter remains \emph{coupled} to the Dilaton \cite{Rahman:2022jsf} and behaves like the $s$-wave sector of higher dimensional matter. Additionally, \cite{Maldacena:2019cbz,Cotler:2019nbi} focus on the case of a timelike dilaton which functions as a clock, while \cite{Sybesma:2020fxg,Svesko:2022txo,Rahman:2022jsf} focus on the case of a spacelike dilaton which encodes the entropy (as we do here).} and in \dk . By the ``semiclassical limit" we mean a number of equivalent things:

		\begin{enumerate}
			\item The ratio of the de Sitter radius $\ell_{\mathrm{dS}}$ and the  length $G$ goes to $\infty.$
			\item The de Sitter entropy $S$ goes to $\infty$
			\item In the DSSYK dual the number of fermion modes $N$ goes to $\infty$
		\end{enumerate}
		
			\subsection*{Aside}
		As we will discuss in Appendix \ref{semiclassical}, two distinct things are meant by ``the semiclassical limit" in quantum gravity---the ``strong" semiclassical limit and the ``weak" semiclassical limit. Both limits assume the above. 
	\subsection*{End of Aside}
		
		\quad In the limit described above there is a ``separation of scales": the natural energy/length scales for various relevant physical phenomena may differ from one another by factors which go to $0$ or $\infty$ \cite{Susskind:2022bia}. Dimensional analysis is subtle precisely because the factors relating different scales/units---e.g. cosmological and string---diverge.
		
		We will begin by going over the various scales that will be important for what follows, which are illustrated in figure \ref{scales} below.
		\begin{figure}[H]
			\begin{center}
				\includegraphics[scale=.7]{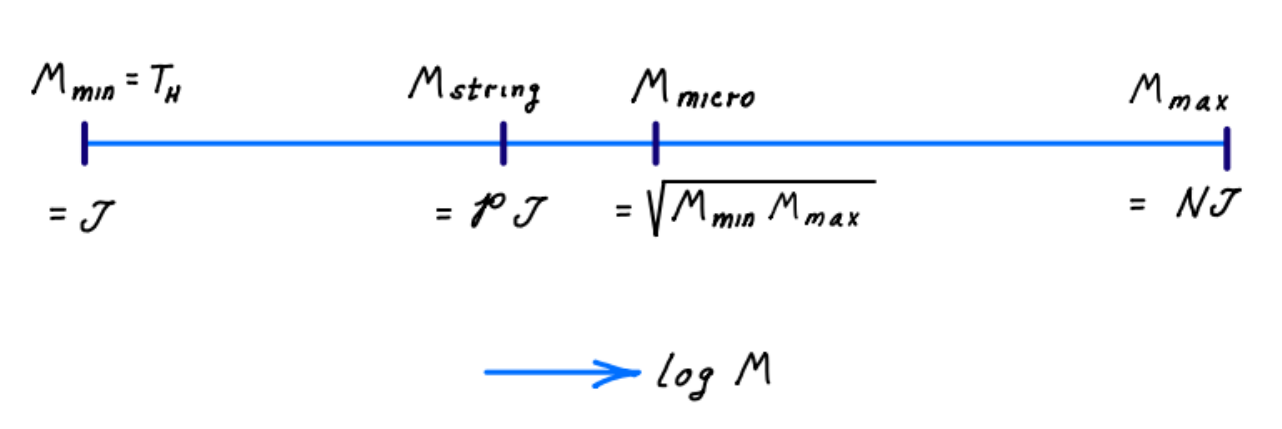}
				\caption{The various scales that appear in \dk/dS$_3$ plotted logarithmically.}
				\label{scales}
			\end{center}
		\end{figure}
		The lowest mass scale in dS$_3$ (i.e. the inverse of the largest Compton wavelength that can fit inside the static patch) is given by $M_{\mathrm{min}} = T_H \sim 1/\ell_{\mathrm{dS}}$ with $T_H$ the Hawking temperature
		\begin{equation}
			\boxed{T_{H} = \frac{1}{2\pi\ell_{\mathrm{dS}}}}
		\end{equation}
		In the \dk \ dual it is the energy parameter $\CJ$ \cite{Lin:2022nss} \emph{provided both quantities are measured in the same units} (we will explain clearly what is meant by this, as well as the argument for the identification $T_H \sim \mathcal{J}$ in the following two sections).
		
	The maximum mass that can exist in dS$_3$ is $M_{\mathrm{max}} \sim 1/G$. In the dual theory $M_{\mathrm{max}}$ goes as $N\mathcal{J}$, which follows from the identification $T_H \sim \mathcal{J}$ and our central assumption $S_{\mathrm{dS}} \sim N$ after noting that, from general relativity,
		\begin{equation}
			\boxed{M_{\mathrm{max}} \ \sim \ S_{\mathrm{dS}}T_H \ \underset{\mathrm{holography}}{\sim} \ N\mathcal{J}}
            \label{maxPrim}
		\end{equation}
        The meaning of $M_{\mathrm{max}}$ is that it is the mass which corresponds to a maximal conical deficit angle of $2\pi$ in $(2+1)$-dimensional de Sitter space  and which represents a minimum entropy state of the static patch (see Appendix \ref{massApp}).
		Equation \eqref{maxPrim} will play a prominent role in analyzing the discrepancy with \cite{Narovlansky:2023lfz} where NV argue for a very different relation,
		\be 
		M_{\mathrm{max}} \ \sim \ \frac{\mathcal{J}}{\lambda}
		\label{MJL}
		\ee
	We remind the reader that $\lambda = 2p^2/N$ is kept finite and $O(1)$ in the double-scaled limit.
  
		The micro scale is defined as the geometric mean of these scales, given by  $M_{\mathrm{micro}} \sim \sqrt{N}\mathcal{J}$, and the string scale  is typically somewhat lower (by a factor of $\sqrt{\lambda}$), at $M_{\mathrm{string}} \sim p\mathcal{J}$ \cite{Susskind:2022bia}. The meaning of the micro scale is that it is precisely the scale at which we can ignore both gravitational backreaction and cosmological curvature and therefore approximate physics by nongravitational flat space physics. We will discuss the meaning of the string scale below.
		
	To summarize, we have
		\begin{equation}
			\boxed{M_{\mathrm{min}} \ = \ T_{H} 
				\ \underset{\mathrm{holography}}{\sim} \ \mathcal{J}}
			\label{min}
		\end{equation}
		\begin{equation}
			\boxed{M_{\mathrm{max}} \ \sim \ 1/G \ \underset{\mathrm{holography}}{\sim} \  N \mathcal{J}}
			\label{max}
		\end{equation}
		\begin{equation}
			\boxed{M_{\mathrm{micro}} = \sqrt{M_{\mathrm{min}}M_{\mathrm{max}}} \ \underset{\mathrm{holography}}{\sim} \ \sqrt{N}\mathcal{J}}
			\label{micro}
		\end{equation}
		\begin{equation}
			\boxed{M_{\mathrm{string}} \ \underset{\mathrm{holography}}{\sim} \ p\mathcal{J}}
			\label{string}
		\end{equation}
		We emphasize again that all of the preceding require both sides to be measured in the same units.
		
		Length scales inverse to the mass scales can be defined via
		\be 
		\boxed{l _{\mathrm{min}}
			= 1/M_{\mathrm{min}} \ \sim \ \ell_{\mathrm{dS}}}
		\label{l}
		\ee
		\be 
		\boxed{l_{\mathrm{string}} = 1/M_{\mathrm{string}}}
		\label{lstring}
		\ee
		and
		\be 
		\boxed{l_{\mathrm{micro}} = 1/M_{\mathrm{micro}}}
		\label{lmicro}
		\ee
        Since $M_{\mathrm{min}} = T_H \ \sim \ 1/\ell_{\mathrm{dS}}$, we will sometimes refer to  $M_{\mathrm{min}}$  and $l_{\mathrm{min}}$ as  \emph{cosmic scales}.
		\begin{equation}
			\boxed{M_{\mathrm{cosmic}} \ \equiv \ M_{\mathrm{min}}}
		\end{equation}
		\begin{equation}
       \boxed{l_{\mathrm{cosmic}} \ \equiv \ l_{\mathrm{min}} \ \sim \ \ell_{\mathrm{dS}}}
		\end{equation}

		\subsection*{Aside: The Role of the Planck Scale in 4D vs 3D}
		\quad The Planck scale plays two independent roles in $(3 + 1)$-dimensional de Sitter space. The first is the more familiar one, namely it acts as an entropy counting parameter that normalizes the horizon area in the relationship 
		\begin{equation}
			\boxed{S_{\mathrm{GH}} = \frac{\mathrm{Area}}{4G}}
		\end{equation}
		The other is that the Planck mass\footnote{In our world the Planck mass is aproximately the mass of a dust particle.} 
		\begin{equation}
			\boxed{M_{\mathrm{Planck}} \ = \ G^{-\frac{1}{D-2}}}
		\end{equation}
		is---in 4D and only in 4D---precisely the mass scale at which one can ignore both gravitational backreaction and cosmological curvature and hence approximate physics by nongravitational flat space physics. This latter role follows from the fact that it---again, in 4D and only in 4D---the Planck mass is \emph{also} given by the geometric mean 
		\begin{equation}
			\boxed{M_{\mathrm{Planck}} \ \underset{\text{4D only}}{=} \ \sqrt{T_HM_{\mathrm{Nariai}}}}
		\end{equation}
		of the minimum (again $T_H$) and maximum (now the Nariai mass $M_{\mathrm{Nariai}}$) energy scales in dS$_4$. This second definition of the Planck mass is directly analogous to the definition \eqref{micro} of the ``micro scale" given above, i.e. we have 
		\begin{equation}
			\boxed{M_{\mathrm{Planck}} \ \underset{\text{4D only}}{=} \ M_{\mathrm{micro}}}
		\end{equation}
   This latter role is taken over in general dimension by the micro scale $M_{\mathrm{micro}}$ defined in \eqref{micro} above. To reemphasize, in 3D it is the micro scale which plays the role of the 4D Planck \emph{mass}, while it is the max scale
			\begin{equation}
				\boxed{l_{\mathrm{max}} = 1/M_{\mathrm{max}} \ \sim \ G}
			\end{equation}
		which plays the role of the Planck \emph{length} in the sense of being an entropy counting parameter 
			\begin{equation}
				\boxed{S_{\mathrm{dS}} \ \sim \ \frac{\ell_{\mathrm{dS}}}{G} \ \sim \ \frac{l_{\mathrm{cosmic}}}{l_{\mathrm{max}}}}
		\end{equation}

          \subsection*{Chord Operators and the Meaning of the ``String Scale"}
          \quad In \dk \ a scale occurs which is associated with so-called ``chord operators" \cite{Berkooz:2018jqr, Lin:2022rbf, Lin:2023trc}. A chord is a collection of $\sim p$ fundamental fermions with corresponding mass scale $\sim p \CJ$. This is precisely what we have called the ``string scale" above (compare with \eqref{string}. 

          Why do we call this scale the string scale? The reason is that the scale $M_{\mathrm{string}}$ plays a role analogous to that of the confinement scale---the scale at which hadrons form string-like flux tubes---in QCD. To understand this role, let's recall ordinary perturbation theory in QCD as well as in DSSYK$_{\infty}$. The perturbation expansions are extremely similar \cite{Susskind:2023hnj}. In particular, both expansions are infrared (IR) divergent: every Lorentz-signature Feynman diagram diverges at large time even if space is compact. In Euclidean signature these IR divergences are regulated (for an already compact spatial geometry), but only because of the nonzero temperature, which makes the Euclidean time direction compact as well. However, the real IR regulator in QCD is not the finiteness of the Euclidean background, but rather it is the confinement/string scale, which is only visible after resumming an infinite number of diagrams. As a length scale, this scale is typically much smaller than the scale of the (assumed compact) spatial geometry. Exactly the same thing is true in DSSYK$_{\infty}$. 
          
          In QCD we must sum the infinite number of planar diagrams (genus zero ribbon diagrams) which have a well-known relation \cite{tHooft:1973alw} to string theory worldsheet diagrams. In DSSYK$_{\infty}$ we must sum the infinite number of melon diagrams (see e.g. \cite{Maldacena:2016hyu}). For example, the lowest order perturbative self energy diagram for a chord two-point function involves an integrand which is time independent and therefore diverges. But summing the infinite number of melon diagrams, we find that (at least for small\footnote{There are corrections to \eqref{CC} for finite $\lambda$ and finite $p$ (as well as, of course, for finite $N$).} $\lambda$) the chord two-point function actually behaves like the integrable function
          \begin{equation}
              \langle\,\mathrm{Chord}(t)\,\mathrm{Chord}(0)\,\rangle = \frac{1}{\cosh^2(p\mathcal{J}t)}
              \label{CC}
          \end{equation}
          with an emergent IR energy scale $\sim p\mathcal{J}$. In writing \eqref{CC} we have employed the SYK conventions described in Section \ref{DimAnal} below, see \eqref{HDSSYK}, \eqref{J2}. The parallel between the sum of the melon diagrams in DSSYK and the planar diagrams of QCD is the reason why $p\mathcal{J}$ was originally called the string scale in our earlier works.
          
          In the context of the DSSYK$_{\infty}$/dS duality conjectured by RS, the chord/string scale plays a role analogous to that of the usual string scale in $(3 + 1)$ dimensions. Namely, it is a measure of bulk (non)locality: the smaller $l_{\mathrm{string}}$, the more local the theory. When the string scale is of order the micro scale\footnote{This is like saying when the string scale is of order the Planck scale \emph{in (3 + 1) dimensions}.}, it tends to zero in cosmic units (see \eqref{semi} below). If chord/string non-locality is the main source of non-locality, the theory then becomes ``sub-cosmically" local as $l_{\mathrm{string}}/l_{\mathrm{dS}}\to 0$. ``Sub-cosmic locality" is the analog of the sub-AdS locality which takes place in the flat-space limit of AdS/CFT.

        \subsection*{End Asides}
        
	\quad The relations between the scales is summarized by,
		
		\begin{equation}
			\boxed{\frac{l_{\mathrm{string}}}{l_{\mathrm{cosmic}}} \ \underset{\mathrm{holography}}{\sim} \ \frac{1}{p} \ }
			\label{stringy}
		\end{equation}
		\begin{equation}
			\boxed{\frac{l_{\mathrm{micro}}}{l_{\mathrm{cosmic}}} \ \underset{\mathrm{holography}}{\sim} \ \frac{1}{\sqrt{N}} \ }
			\label{semi}
		\end{equation}
		\begin{equation}
			\boxed{\frac{l_{\mathrm{micro}}}{l_{\mathrm{string}}} \ \underset{\mathrm{holography}}{\sim} \ \sqrt{\lambda} \ }
			\label{microstring}
		\end{equation}
		where $\lambda = 2p^2/N$ is kept finite in the double-scaled limit. The relationship
		\eqref{microstring} in the form 
		\begin{equation}
			\boxed{\inp{\frac{l_{\mathrm{micro}}}{l_{\mathrm{string}}}}^2 = \lambda}
		\end{equation}
		demonstates that the sub-cosmically local limit (in the sense described above) corresponds to the double-scaled limit in which 
 $\lambda$ is finite, $O(1)$, and nonzero. 
		
		\subsection{Units and Dimensions in the Semiclassical Limit}
		\label{units}
		\quad Dimensional analysis in the semiclassical limit is subtle, and it is very important to be clear about it. Let us consider a certain quantity $X$ with dimensions of length, for example the distance between Palo Alto and San Francisco, which is about 50 kilometers. One way of thinking about $X$ is ``conceptual": just what we said---the distance from Palo Alto to San Francisco. For that we don't need to specify any choice of units. But to give $X$ a numerical value we do need units:
		\begin{align}
			X  
			\ &\approx \ 50\,\rm{km} \\
			\ &\approx \ 4\times10^{-22}\,\text{Hubble radii} \\
			\ &\approx \ 3\times 10^{39}\,\text{Planck lengths}
			\label{XPl}
		\end{align}
		
		In what follows, we will find it helpful to introduce the notation $[X]_{unit}$ to denote the \emph{numerical value} of the dimensionful quantity $X$ in the units of $unit$, i.e.
		\begin{align}
			[X]_{\mathrm{kilometers}} \ &\approx \ 50 \\
			[X]_{\mathrm{Hubble}} \ &\approx \ 4\times10^{-22} \\
			[X]_{\mathrm{Planck}} \ &\approx \ 3\times 10^{39}
		\end{align}

		 Things become  more subtle when comparing units corresponding to scales that separate in some limit\footnote{Two scales are said to separate when their ratio diverges or goes to zero in some limit.}. For example, in cosmic units  the de Sitter radius $\ell_{\mathrm{dS}}$ is
		\begin{equation}
			[\ell_{\mathrm{dS}}]_{\mathrm{cosmic}} = \frac{1}{2\pi}
		\end{equation}
		In string units, this same quantity has numerical value
		\begin{equation}
			[\ell_{\mathrm{dS}}]_{\mathrm{string}} = \frac{p}{2\pi} \ \to \ \infty 
		\end{equation}
		and in micro units it has numerical value
		\begin{equation}
			[\ell_{\mathrm{dS}}]_{\mathrm{micro}} = \frac{\sqrt{N}}{2\pi} \ \to \ \infty
		\end{equation}
		
		We usually write equations in dimensionally consistent form so that they are true in any unit system. An example would be the relationship between the distance $Y$ from San Francisco to New York City and the distance $X$ from San Francisco to Palo Alto:
		\begin{equation}
			Y \ \approx \ 87\times X.
			\label{YX}
		\end{equation}
		 It is of course important to make sure that the units on both sides of \eqref{YX} are the same\footnote{  We are purposely being very pedantic in emphasizing these points because we have found in discussions with colleagues that confusions arise  from conflating the numerical values of dimensionful quantities in different units.}, since
		\begin{equation}
			[Y]_{\mathrm{miles}} \ \neq \ 87\times [X]_{\mathrm{inches}}
		\end{equation}
		Similarly for the Hawking temperature, 
		\begin{equation}
			\boxed{[T_H]_{\mathrm{string}} \neq [T_H]_{\mathrm{cosmic}}}
		\end{equation}

		We can now be clear about what we meant by the equations of Section \ref{SepScales} above: we meant that 
		\begin{equation}
			\boxed{[M_{\mathrm{min}}]_{\mathrm{cosmic}} \ = \ [T_{H}]_{\mathrm{cosmic}} 
				\ \underset{\mathrm{holography}}{\sim} \ [\mathcal{J}]_{\mathrm{cosmic}}}
		\end{equation}
		etc. 
  
  Let's apply these ideas of the separation of scales and the corresponding dimensional analysis to the double-scaled SYK model. In particular,  we will explain how accounting for the various possible units/scales (i.e. cosmic, string, and micro) will  clarify and unify the various different conventions which exist in the literature.
		
		\subsection{Dimensional Analysis in DSSYK}
		\label{DimAnal}
		\quad The Hamiltonian of the $p$-local SYK model has the form 
		
		\begin{equation}
			\boxed{H \ = \mathrm{i}^{p/2}\,\sum_{i_1 < i_2 < \,\dots\, < i_p}J_{i_1i_2\,\dots\, i_p}\,\psi_{i_1}\psi_{i_2}\dots\psi_{i_p}}
			\label{HDSSYK}
		\end{equation}
		with random couplings $J_{i_1\dots i_p}$ distributed according to a Gaussian of mean zero and of variance\footnote{More precisely, we have
			\begin{equation}
				\langle J_{i_1\dots i_p}J_{j_1\dots j_p}\rangle = \langle J^2 \rangle \,\delta_{i_1j_1}\dots\delta_{i_pj_p}
			\end{equation}
			We have chosen \eqref{J2} in such a way as to always reproduce (up to factors of $2$) the correct formula for the variance once the energy units/scale for the Hamiltonian and energy parameter $\mathcal{J}$ have been specified, see \eqref{J2c}-\eqref{J2m}.
		}
		\be
		\boxed{  
			\la J^2   \ra \ \equiv \ \frac{\mathcal{J}^2N}{\binom{N}{p}} \ \sim \ \frac{\mathcal{J}^2p!}{N^{p-1}}}
		\label{J2}
		\ee
		
		Equation \eqref{J2} hides a subtlety. The Hamiltonian, having dimensions of energy or inverse length, has values which depend on the choice of units/scale, as we have explained at length above.  We can account for unit-dependence  by putting subscripts on dimensionful quantities---for example 
		writing $[H]_{x}$ where $x$ can mean cosmic, string, or micro. 
		The same can be said for the  dimensionful parameters $J$ and $\CJ$ which also have units of energy. Since the units on both sides of \eqref{J2c} match, the equation is true in any units, but the numerical values of energy, $J$, and  $\CJ$ do depend on the choice of units.
  
		The numerical 
		values of $\CJ$ in the three unit/scale systems satisfy
		\begin{equation}
			\boxed{
				[\CJ]_{\mathrm{cosmic}} \equiv \CJ_0}
			\label{Jc}
		\end{equation}
		\begin{equation}
			\boxed{
				[\CJ]_{\mathrm{string}} \ \sim \ \frac{\CJ_0}{p}  }
			\label{Jst}
		\end{equation}
		\begin{equation}
			\boxed{[\CJ]_{\mathrm{micro}} \ \sim \  \frac{\CJ_0}{\sqrt{N}}  }
			\label{Jm}
		\end{equation}
		Here we have \emph{defined} $\mathcal{J}_0$ to agree with the \emph{numerical value} of $\mathcal{J}$ \emph{in cosmic units}. This is a \emph{fixed quantity} (which will of course depend on our choice of cosmic rods and clocks) which is assumed to be $O(1)$. \eqref{Jst} and \eqref{Jm} then follow from \eqref{l}-\eqref{microstring}. 
		
		We can now write the Hamiltonian in the three unit systems as
		\begin{equation}
			\boxed{	[H]_{x} \ = \sum_{i_1 < i_2 < \,\dots\, < i_p}  [J_{i_1i_2\,\dots\, i_p}]_x\,\psi_{i_1}\psi_{i_2}\dots\psi_{i_p} }
			\label{Hx}
		\end{equation}
		with the variance in $J$ depending on the choice of units/scale---i.e. depending on the value of $x$: 
		\be
		\boxed{  
			\la J^2   \ra_x = \frac{N}{\binom{N}{p}}\,[\mathcal{J}]_x^2 \ \sim \ \frac{p!}{N^{p-1}}\,[\mathcal{J}]_x^2}
		\label{J2x}
		\ee
		Plugging in the definitions \eqref{Jc}, \eqref{Jst}, \eqref{Jm}, we find, respectively, that 
		\begin{equation}
			\boxed{  
				\langle J^2 \rangle_{\mathrm{cosmic}} = \frac{N}{\binom{N}{p}}\,\mathcal{J}_0^2 \ \sim \ \frac{p!}{N^{p-1}}\,\mathcal{J}_0^2}
			\label{J2c}
		\end{equation}
		\begin{equation}
			\boxed{  
				\langle J^2 \rangle_{\mathrm{string}} = \frac{N}{p^2\binom{N}{p}}\,\mathcal{J}_0^2 \ \sim \ \frac{p!}{p^2N^{p-1}}\,\mathcal{J}_0^2}
			\label{J2s}
		\end{equation}
		\begin{equation}
			\boxed{  
				\langle J^2 \rangle_{\mathrm{micro}} = \frac{1}{\binom{N}{p}}\,\mathcal{J}_0^2 \ \sim \ \frac{p!}{N^{p}}\,\mathcal{J}_0^2}
			\label{J2m}
		\end{equation}
		
		Equations \eqref{J2c}\eqref{J2s}\eqref{J2m}  correspond (up to possible factors of $2$) to the three conventions commonly used in the DSSYK literature. Equation \eqref{J2c} is the convention used by the authors (RS) in \cite{Susskind:2021esx,Susskind:2022dfz,Susskind:2022bia,Rahman:2022jsf} as well as in \cite{Lin:2022nss}, where the emphasis is almost entirely on cosmic scales. Maldacena and Stanford \cite{Maldacena:2016hyu} use the ``string" convention \eqref{J2s} which is also used by NV \cite{Narovlansky:2023lfz} and by Lin and Stanford \cite{Lin:2023trc}. 
	Berkooz and collaborators \cite{Berkooz:2018jqr} use the convention \eqref{J2m} corresponding to the use of micro-scale units.  
		
\section{Three Temperatures}		
\label{BvH}
\quad There are three distinct concepts of temperature that appear in the holographic formulation of de Sitter space. These seem to be different and not related by just a change of units. The first is the ``Boltzmann temperature" $T_B$ which is the temperature parameter that appears in the thermal density matrix,
\be \boxed{
\rho = \frac{1}{Z}\,\exp{\inp{-H/T_B}}}
\label{density}
\ee
What we know about $T_B$ is that it is infinite in cosmic units
\be \boxed{ 
[T_B]_{\mathrm{cosmic}} = \infty}
\label{Tb=infty}
\ee
(see for example \cite{Susskind:2021omt} as well as separate arguments by Banks \cite{Banks:2003cg, Banks:2006rx}, Fischler \cite{Fischler}, Dong et al \cite{Dong:2018cuv}, and Chandrasekaran et al \cite{Chandrasekaran:2022cip}). Indeed the $\infty$ in \dk \ is meant to refer to the value of $T_B$.

We cannot conclude from this that the Boltzmann temperature is also infinite in string units since the ratio of the string scale to the cosmic scale goes itself to $\infty$ in the double-scaled limit.
Indeed there are reasons to believe that \eqref{Tb=infty} should be refined to read,
\be
[T_B]_{\mathrm{cosmic}} \ \sim \ p\CJ_0 \ \to \ \infty  \quad \text{(in double-scaled limit)}
\label{Tb=J0p}
\ee
This would change nothing in the analysis of cosmic-scale phenomena but can affect $1/N$ corrections to string-scale phenomena. Changing  to string units in \eqref{Tb=J0p}  gives,
\be 
[T_{B}]_{\mathrm{string}} \ \sim \ \CJ_0 \qquad \text{(speculative)}
\label{Tbstring}
\ee
Here we see an interesting point: infinite temperature in cosmic units does not necessarily mean infinite temperature in string units. The infinity in \dk \ should always be interpreted as infinite temperature in cosmic units.
For now this is simply an aside, but we will return to this point in a future publication.


The second concept of temperature is the emergent Hawking temperature $T_H$ which should be identified with the notion of ``Tomperature" defined in   \cite{Lin:2022nss}. Let us briefly explain what is meant by this. The Hawking temperature\footnote{The Hawking temperature is also the proper temperature of fluctuations as seen from the pode \cite{Gibbons:1977mu}.} $T_H$ is defined by the probability for a small fluctuation of bulk mass $M$ \cite{Gibbons:1977mu,Susskind:2021omt,Chandrasekaran:2022cip}
\be \boxed{
P_{\mathrm{fluct}} \ \sim \  e^{-\Delta S} \equiv e^{-M/T_H}}
\label{Prob}
\ee
Here $\Delta S$ is the entropy deficit associated with the fluctuation. In the bulk, $\Delta S$ is simply the change in the horizon area (i.e. in the Gibbons-Hawking entropy)
. We must be very careful in how we identify $\Delta S$ in the dual quantum theory.

The $\Delta S$ that appears in \eqref{Prob} is \emph{not} the $\delta S$ that appears in the usual first law of thermodynamics of the dual quantum theory
\begin{equation}
    \boxed{\delta S = \delta E/T_B}
    \label{first}
\end{equation}
Equation \eqref{first} is defined by studying quasistatic processes involving small energy changes or, equivalently, by looking at the slope of a fixed energy distribution as a function of the Von-Neumann entropy as we slightly change the state. In particular, when using \eqref{first} one must keep all parameters of the system fixed and one will find that the temperature determined in this way is the 
Boltzmann temperature discussed above. The definition of Tomperature by contrast is meant to holographically capture the physical process of Hawking emission, in which one degree of freedom\footnote{Really two Fermionic degrees of freedom, which is the same as one qubit in the standard Hilbert space representation of the SYK model. See \cite{Lin:2022nss} for details.} is ``emitted"---i.e. frozen out from---the horizon while leaving the couplings involving all \emph{other} degrees of freedom fixed. With $\Delta S$ defined in this way, one finds that \cite{Lin:2022nss}
\begin{equation}
    \boxed{\Delta S = \Delta E / (2\mathcal{J})}
\end{equation}
where we emphasize that the above is true when both $E$ and $\mathcal{J}$ are written in the same units/measured at the same scale. It is this quantity 
\begin{equation}
    \boxed{\mathrm{Tomperature} \ = \ 2\mathcal{J}}
\end{equation}
which will agree with the Hawking temperature defined by \eqref{Prob} above. (Here we are implicitly using the fact, which we will explain in more detail below, that bulk mass and quantum mechanical energy---as defined by the DSSYK Hamiltonian \eqref{HDSSYK}---should agree when both are near zero). We therefore find that $T_H$ satisfies the unit-independent relation,
\be \boxed{
T_H \ \sim \ \CJ}
\label{T_H=J}
\ee
or
\be \boxed{
[T_H]_x  \ \sim \ [\CJ]_x } 
\label{THx=Jx}
\ee
The Hawking temperature/Tomperature is finite in cosmic units and transforms when we change units. In particular, we have that 
\be \boxed{
[T_H]_{\mathrm{cosmic}}  \ \sim \ \CJ_0 }
\ee
and 
\be \boxed{
[T_H]_{\mathrm{string}} \ \sim \ \frac{\mathcal{J}_0}{p} }
\ee

A third concept of temperature is the temperature $T_{\mathrm{chord}}$ seen by chords. To understand $T_{\mathrm{chord}}$ we need to know what a chord is. We will not get into the precise definition of chords (see e.g. \cite{Berkooz:2018jqr, Lin:2022rbf, Lin:2023trc} for this) but will simply note as we did above that chords are collections of $\sim p$ fermions which are confined to the stretched horizon region\footnote{This is in agreement with the ``fake disk" picture of Lin and Stanford \cite{Lin:2023trc}. We thank Henry Lin for discussions on this point as well as for suggesting the following calculation.} \cite{Susskind:2023hnj}. They therefore live in a hot environment \cite{Susskind:2023hnj} where the proper temperature that they experience is given by blue-shifting\footnote{Note however that the Hamiltonian \eqref{HDSSYK} of the whole system still remains conjugate to \emph{Rindler time}, rather than proper time. Indeed, bulk excitations which make it out to the pode will experience the finite Hawking temperature $T_H$. It is only because the chords are confined to the near horizon region that they \emph{experience} the blueshifted temperature $p\,T_H$.} the coordinate/pode Hawking temperature. The blue-shift factor is
\be \boxed{
\text{blue shift} = \sqrt{\frac{g_{tt}(r=0)}{g_{tt}(r_{\mathrm{sh}})}} 
}
\label{blueshift}
\ee
Here $g_{tt}$ is the time-time component of the metric and $r_{\mathrm{sh}}$ means the value of the radial coordinate at the stretched horizon. Assuming that the stretched horizon is at a proper distance $\sim O\inp{l_{\mathrm{string}}}$ from the bifurcate horizon\footnote{In previous works \cite{Rahman:2022jsf} we had guessed that this distance might be of order a Planck/max length, but we now believe, following \cite{Susskind:2023hnj}, that this distance should be of order a string length.}, and using the standard de Sitter metric (see e.g. \eqref{SdS}), we find that\footnote{Specifically, we will find that
\be
\text{proper distance} \ \sim \ l_{\mathrm{string}} \ \implies \ \frac{r_{\mathrm{sh}}}{\ell_{\mathrm{dS}}} = 1-\frac{1}{2}\inp{\frac{l_{\mathrm{string}}}{\ell_{\mathrm{dS}}}}^2 + O\big(l_{\mathrm{string}}^4/\ell_{\mathrm{dS}}^4\big)
\label{rsh}
\ee
giving 
\begin{equation}
    g_{tt}(r_{\mathrm{sh}}) = 1 - \frac{r_{\mathrm{sh}}^2}{\ell_{\mathrm{dS}}^2} \ \sim \ \inp{\frac{l_{\mathrm{string}}}{\ell_{\mathrm{dS}}}}^2 \ \sim \ \frac{1}{p^2}
\end{equation}
finally giving
\begin{equation}
    \text{blue shift} = \sqrt{\frac{g_{tt}(r=0)}{g_{tt}(r_{\mathrm{sh}})}} \ \sim \ p
\end{equation}
as claimed.} 
\be 
\boxed{\text{blue shift} \ \sim \ p}
\label{blueshift=p}
\ee
so that the temperature experienced by chords is,
\be 
\boxed{[T_{\mathrm{chord}}]_{\mathrm{string}} = \CJ_0}
\label{Tch=J}
\ee
Indeed, comparing with \eqref{Tbstring} we see that this is the same as the Boltzmann temperature in string units that was conjectured above.

The fact that chords live in a hot environment with temperature \eqref{Tch=J} explains a number of things which we list here:
\begin{enumerate}
    \item Chord correlations, which have the form 
    \be 
[\la \CO(t)\,\CO(0) \ra]_{\mathrm{string}} \ \sim \ \frac{1}{\cosh^2(\CJ_0 \, [t]_{\mathrm{string}})}
\label{chch}
    \ee
    are periodic with period $\pi / \CJ_0$ when continued to Euclidean signature. This is precisely\footnote{Note that the inverse circumference of the fake disk disagrees with the Tomperature in string units $[\mathrm{Tomperature}]_{\mathrm{string}} \sim \mathcal{J}_0/p$ by a factor of $1/p$. But this factor is precisely accounted for by the blueshift factor considered above. $\mathcal{J}_0/p$ is the \emph{coordinate} Hawking temperature in string units, or, equivalently, the proper Hawking temperature in string units \emph{as seen at the pode}. But it is $$\mathcal{J}_0 = [T_{\mathrm{chord}}]_{\mathrm{string}} \ \sim \ p\times [\mathrm{Tomperature}]_{\mathrm{string}}$$
    which is \emph{felt} by the Chords, which are confined to the hot near-horizon region. We thank Henry Lin for discussions on this point.} the circumference of the ``fake disk" discussed by Lin and Stanford \cite{Lin:2023trc}.
    
    \item Chord correlations factorize into products of single fermion correlations for small $\lambda$. This simply reflects the fact that in a hot environment like a plasma, composite objects just fall apart into their constituents.
    
    \item Chord correlations decay exponentially with time, which is a standard property of objects propagating through a plasma.
\end{enumerate}

For our present purpose of responding to NV, the main point that we wish to emphasize is that, in string units, $\CJ_0$ is not the Hawking temperature appropriate to observations at the pode, but is rather the blue-shifted temperature seen by confined chords. The point is that, in string units, we should have 
\begin{equation}
    \boxed{[T_H]_{\mathrm{string}} = \frac{\mathcal{J}_0}{p}}
\end{equation}

		\section{The Argument of NV}

		\quad Let's begin by explaining the argument made by NV in \cite{Narovlansky:2023lfz} that leads to the formula \eqref{NV0}
		\begin{equation}
			\frac{\ell_{\mathrm{dS}}}{G} \ \underset{\text{claimed by NV}}{\sim} \ \frac{1}{\lambda}
			\label{NV00}
		\end{equation}
		(we remind the reader that we consider only boxed formulas  to be correct).
		
	NV begin by citing the energy spectrum of DSSYK found by \cite{Berkooz:2018qkz}, which can be parameterized by an angle 
	$\theta \in [0,\pi]$
	\footnote{Recall that, as we have explained above, NV work in what we have 
called string units and also that what we call $\CJ_0$ they call (up to $\lambda$-independent $O(1)$ factors) $\mathbb{J}$. }   
\begin{equation}
			\boxed{[E(\theta)]_{\mathrm{string}} \ \sim \ -\frac{\CJ_0   \cos(\theta)}{\lambda}}
			\label{Etheta}
		\end{equation}
		(Eq. (4) of reference \cite{Narovlansky:2023lfz}). This implies that the maximum value of the energy in string units is\footnote{Or, more generally, 
			\begin{equation}
				\boxed{[E_{\mathrm{max}}]_{x} \ \sim \ \frac{p}{\lambda}[\CJ_0]_x}
				\label{Emax}
			\end{equation}} 
		\be \boxed{
		[E_{\mathrm{max}}]_{\mathrm{string}} \ \sim \ \frac{\CJ_0}{\lambda} 
		}
		\label{Emaxstring}
		\ee
		
Small deviations from the zero energy state\footnote{NV work in the microcanonical ensemble peaked around the zero energy state rather than in the canonical ensemble at infinite Boltzmann temperature. This does not lead to any significant differences with RS.} can be parameterized as 
		\begin{equation}
			\boxed{[\delta E]_{\mathrm{string}} \ \approx \ \frac{\CJ_0}{\lambda}\,\pi v} 
			\label{dE}
		\end{equation}
		with 
		\begin{equation}
			\boxed{\pi v \equiv \pi - 2\theta \ll 1}
			\label{piv}
		\end{equation}
		(Eq (5) of \cite{Narovlansky:2023lfz}). Note that $\pi v$ is defined to be small \emph{numerically}, not parameterically. In other words we still have that $\pi v \sim O(1)$ parameterically.
		
		NV then correctly say that for small changes in mass $\delta M$ near $M = 0$, the bulk first law of thermodynamics gives
		\begin{equation}
			\boxed{\delta S = \frac{\delta M}{T_H}}
		\end{equation}
		They also correctly say that for small changes in energy $ \delta E$ centered about $E = 0$, that $\boxed{ \delta E = \delta M}$ (see Section \ref{EvM}) so that the above becomes 
		\begin{equation}
			\boxed{\delta S = \frac{\delta E}{T_H}}
		\end{equation}
		These equations are correct provided the numerator and denominator are evaluated in the same units. We can then use \eqref{dE} to write
		\begin{equation}
			\boxed{\delta S \ \sim \ \frac{1}{[T_H]_{\mathrm{string}}}\frac{\CJ_0}{\lambda}\,\pi v}
		\end{equation}

		NV then claim that 
		\begin{equation}
			[T_H]_{\mathrm{string}} \ \sim \ \mathcal{J}_0
			\label{E1}
		\end{equation}
		to find that 
		\begin{equation}
			\delta S \ \sim \ \frac{1}{\lambda}\,\pi v
		\end{equation}
		(Equation (9) of reference \cite{Narovlansky:2023lfz}).

		NV then also correctly identify that in the bulk, a change in the entropy can be accomplished by introducing a conical defect (i.e. a localized mass, see Appendix \ref{massApp}) of deficit angle\footnote{Here we have used NV's normalization of the conical deficit angle, which differs from the one we use below and denote by $\upalpha$ via \begin{equation}
				\boxed{2\pi\alpha = \upalpha}
		\end{equation}} $2\pi\alpha$, with the corresponding entropy deficit given by 
		\begin{equation}
			\boxed{\delta S = S_{\mathrm{dS}}\,\alpha}
			\label{dS=Sa}
		\end{equation}
		They then conjecture that 
		\begin{equation}
			\pi v = \alpha 
			\label{E2}
		\end{equation}
		which forces  them to the conclusion that 
		\begin{equation}
			S_{\mathrm{dS}} \ \sim \ \frac{1}{\lambda}
			\label{NV9}
		\end{equation}
		(equivalent to eq. (11) of \cite{Narovlansky:2023lfz}).
		
		Let's now state the two points of disagreement with NV: The first point is that 
		\eqref{E1} should really read 
		\begin{equation}
			\boxed{[T_H]_{\mathrm{string}} \ \sim \ \frac{\CJ_0}{p}} 
		\label{TS=J/p}
		\end{equation}
		The second point of diagreement is that we believe that for small deviations from equilibrium the basic principles of holography  require that
		\begin{equation}
			\alpha \ \sim \ \frac{v}{p}
		\end{equation}
		Taking these two points of disagreement into account would then correct \eqref{NV9}  to 
		to
		\begin{equation}
			\boxed{S_{\mathrm{dS}} \ \sim \ \frac{p\CJ_0/\lambda}{\CJ_0/p} \ \sim \ N}
			\label{S=N}
		\end{equation}
		
		Let's now discuss these two points of disagreement in  detail.
		
		\section{First Disagreement}
		\label{Temp}

		\quad The first discrepancy between ourselves and NV  involves the value of the Hawking temperature; namely its value in string vs cosmic units. NV write in string units (see equation (7) of \cite{Narovlansky:2023lfz}),
		\be 
		[T_{H}]_{\mathrm{string}} \ \sim \ \CJ_0    
		\label{THNV}
		\ee
  We have explained in equations \eqref{blueshift}\eqref{blueshift=p}\eqref{Tch=J} that $\CJ_0$ is the blue-shifted temperature seen by chords at the stretched horizon; not the Hawking temperature seen at the pode (which is the one appropriate for use in the bulk first law/the above derivation). 
  
  We can also understand this from dimensional analysis.
	We believe that \eqref{THNV}  is incorrect; what we consider to be true (see section \ref{BvH}) is that \cite{Lin:2022nss}
		\begin{equation}
			\boxed{[T_H]_x \ \sim \ \mathcal{J}_x}
			\label{TH}
		\end{equation}
		i.e. that 
		\begin{equation}
			\boxed{[T_H]_{\mathrm{cosmic}} \ \sim \ \mathcal{J}_0}
		\end{equation} 
Therefore by a change  to string units,
\begin{equation}
			\boxed{[T_H]_{\mathrm{string}} \ \sim \ \frac{1}{p}\,[T_H]_{\mathrm{cosmic}}  \ \sim \  \frac{\CJ_0}{p} }
   \label{THString}
		\end{equation}

		Let us emphasize this point: Energies in string and cosmic units are related by a factor of $p$ with energies in string units being the smaller of the two. For example, in our world the Hawking temperature 
		would by definition be order unity in cosmic units; assuming the string scale is near the Planck scale the Hawking temperature is about $10^{-60}$ in string units. Similarly, the Hawking temperature in cosmic units is $\CJ_0$ but in string units it is $\CJ_0/p.$ Since the double-scaled limit involves $p\to \infty$, the misidentification \eqref{THNV} would be a serious error.
		
		Incorporating this correction would change NV's result from \eqref{NV0} 
		 to 
		\begin{equation}
			S_{\mathrm{dS}} \ \sim \ p\times \frac{1}{\lambda} \ \sim \ \frac{N}{p}
			\label{NVC2}
		\end{equation}
		which is still off from the correct formula \eqref{holo} by a factor of $1/p$.

\section{Second Disagreement}
		
\quad The remaining discrepancy between 		\eqref{TS=J/p} and \eqref{S=N} is due to a disagreement about the relation between the parameter $v$ in \eqref{piv} and the conical deficit angle $\alpha$ \eqref{dS=Sa}. NV assume 
\be  
v=\alpha
\label{v=a}.
\ee
	We will derive a very different relation. To be very clear, our derivation assumes the standard relations between bulk area and entropy that NV dispute. It is not our intention to prove that NV are wrong but just to identify where our differences lie.
	
To avoid confusion we will work in cosmic units.	Transforming equation \eqref{dE} to cosmic units introduces a factor of $p.$ It becomes,

\be
\boxed{[E(\theta) ]_{\mathrm{cosmic}}  \ \sim \ -p\,\frac{\CJ_0}{\lambda} \cos{\theta}  \
\approx \ \frac{p\CJ_0}{\lambda}\,\pi v }
\label{E(thet)}
\ee
Now add the standard relation between deficit angle and mass reviewed in Appendix \ref{defects} (see equation \eqref{aMfull} and \eqref{aM}): 
\be 
\boxed{		
		\upalpha \ \sim \ 8\pi GM}
		\label{a=8pG}
\ee
  For small mass, energy, and deficit angle we may assume that mass and energy are the same. Thus \eqref{a=8pG} becomes, 
 \be  \boxed{
 \upalpha = 8\pi GE \ \sim \ [G]_{\mathrm{cosmic}} \,\frac{p\CJ_0}{\lambda}\,\pi v }
 \label{alph(E)}
 \ee	
i.e.
		\be \boxed{ 
	\frac{	\CJ_0   \pi}{\lambda}\,v  \ \sim \  \frac{\upalpha}{G p}
	}
	\label{vsima}
		\ee
From $\lambda =2p^2/N$ and $[G]_{\mathrm{cosmic}} \sim 1/(N\CJ_0)$ (which follows from our central assumption) we find that
		\be
		\boxed{v\sim p\upalpha}
		\label{vsimalp}.
		\ee

The difference between \eqref{v=a} and \eqref{vsimalp} accounts for the discrepancy between \eqref{TS=J/p} and \eqref{S=N}. To see this we use (the bulk/tomperature version of) the first law of thermodynamics for small deviations from the de Sitter state,
\be 
\boxed{  \delta S = \frac{\delta M}{T_H} = \frac{\delta E}{T_H}
 }
 \label{DS=DE/TH}.
\ee 
Using
\begin{equation}
    \boxed{\delta S = \frac{\upalpha}{2\pi}\,S_{\mathrm{dS}} \ \implies \ S_{\mathrm{dS}} \ \sim \ \frac{\delta S}{\upalpha}}
\end{equation}
in addition to \eqref{DS=DE/TH}, $[T_H]_{\mathrm{cosmic}} \sim \mathcal{J}_0$, and \eqref{E(thet)} then gives,
\be \boxed{  
S_{\mathrm{dS}} \ \sim \ \frac{p\CJ_0}{\lambda} \frac{1}{\CJ_0} \frac{\pi v}{\upalpha}
}
\label{v/a}
\ee

If we now follow NV and use $\pi v/\upalpha \sim 1$ we get NV's formula \eqref{NVC2}. But if we instead use $v/\upalpha \sim p$ we get the more standard (and we believe correct) formula 
\be \boxed{
S_{\mathrm{dS}} \ \sim \ N}.
\label{fund}
\ee
		
	The argument can, and perhaps should be, run backward. By assuming \ref{fund}  we can derive $\boxed{\upalpha \sim v/p }.$

\subsection{The Location of States with Backreaction}
\begin{figure}[H]
	\begin{center}
	\includegraphics[scale=.6]{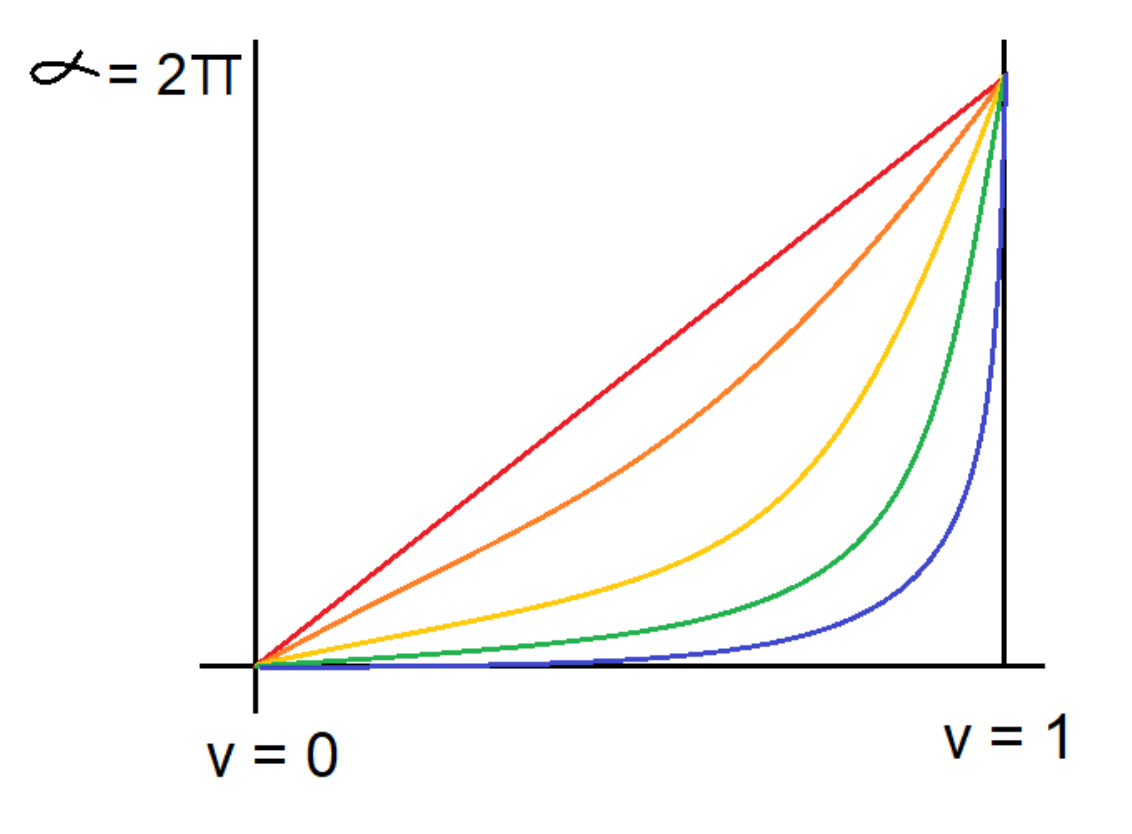}
	\caption{Schematic picture of deficit angle vs $v$ for increasing $p$.}
	\label{av}
	\end{center}
\end{figure}

\quad Let's consider the relation between $v$ and the deficit angle $\alpha.$ First of all we know that when $v=0,$ $\alpha$ also equals $0$. We can extend $v$ away from the regime $v \ll 1$ by simply defining $v \equiv -\cos(\theta)$ so that  when $v$ reaches its maximum at $v=1,$ $\alpha$ reaches its maximum at $\alpha=2\pi$. Secondly, from \eqref{vsima} we have that at small $v$ the relation is linear with slope $1/p,$
\begin{equation}
    \boxed{\upalpha \ \sim \ \frac{v}{p}}
    \label{avp}
\end{equation}
As $p$ increases the slope decreases but in such a way as to preserve the endpoints of the curve. In figure \ref{av} we show  a sequence of curves representing increasing values of $p$ (increasing from red to blue).

We see that for small $v$ the value of $\alpha$ shrinks to zero. This means that for such states back-reaction in the form of a finite deficit angle goes to zero with increasing $p,$ but it does not mean that backreaction in \emph{all} states goes to zero. Instead it means that the states with finite backreaction migrate toward $v=1$, i.e., towards the edge of the energy spectrum, likely into the non-Gaussian tails of the density of states. One does not run out of states since even within these tails the density of states is exponentially large,
\begin{equation}
    \boxed{\text{\# of states at edge} \ \sim \ e^{-\lambda}\,e^N}
\end{equation}

In the double-scaled semiclassical limit $p \sim \sqrt{\lambda N} \to \infty$, states with finite backreaction and appreciable anglular deficits are not found 
near the center of the energy distribution but are swept out to the edges.
This is a manifestation of the separation of scales. Such states with finite deficit angle also have masses of order $M_{\mathrm{max}}$. In the limit they separate from the states with energy in the range $M_{\mathrm{micro}}$ and appear in a delta-function-like spike at $v=1$.

	\subsection*{Aside: Why $M=E$ for Small Deficit Angle}
        \label{EvM}
	\quad Let's consider masses of the order $M_{\mathrm{micro}}$ located near the pode of dS. In the semiclassical limit assuming sub-cosmic locality, the masses effectively move in flat space with negligible backreaction. There are two conditions for this to be true: First the background de Sitter space must have negligible curvature on the micro scale; this is guaranteed by the separation of scales. Secondly the mass must be small enough that it does not create an appreciable deficit angle. If these conditions are satisfied, generators of a local approximate Poincar\'e symmetry can be constructed near the pode which would include spatial momentum as well as energy. Mass and energy can then be defined in the usual way and for a particle at rest near the pode they will coincide, both with each other and with the mass appearing in the metric and the energy determined by the DSSYK Hamiltonian (related to bulk Killing energy) respectively. This is the reason why we may assume $E=M$ for small masses.
	
	But once the mass becomes large enough to make an appreciable deficit angle the approximation of flat space  and of Poincar\'e symmetry breaks down. Mass may still be defined by the parameter in the metric (see appendix \ref{massApp}) but it is no longer connected to symmetry generators. In particular there is no reason to assume it is connected to eigenvalues of the Hamiltonian \eqref{HDSSYK}. When the deficit angle becomes of order $2\pi$ mass and energy can be very different.
	
	Let us consider the extreme situation where the deficit angle is $\upalpha = 2\pi.$ The energy at that point is maximal and from \eqref{E(thet)} we see that it is given by,
	\be  \boxed{
	E_{\mathrm{max}} \ \sim \ \frac{p\CJ}{\lambda}
	}
	\label{Emax}
	\ee
	On the other hand from appendix \ref{massApp} equation \eqref{aMfull} we see that $M_{\mathrm{max}}$ is of order $1/G \sim N\mathcal{J}$. The ratio of the two is therefore given by,
	\be \boxed{
	\frac{M_{\mathrm{max}}}{E_{\mathrm{max}}} \ \sim \ \frac{\lambda N}{p} \ \sim \ p 
	}
	\ee
	
Both $M_{\mathrm{max}}$ and $E_{\mathrm{max}}$ diverge in the \dk \ limit but not in the same way. The ratio goes to infinity as the conical deficit tends to its extreme value of $2\pi.$

		\section{Conclusions}

		\quad This paper is a response to the  recent paper \cite{Narovlansky:2023lfz} of Naravlansky and Verlinde. Despite the fact that we disagree with the conclusions of that paper, we have learned a great deal from it.
		
The first part of \cite{Narovlansky:2023lfz}  arrived at a conclusion \eqref{NV0} regarding the scaling of bulk entropy which disagrees with previous work on the subject \RS as well as with basic holographic expectations. In this note we have traced the   disagreement to two
 key assumptions made by NV. When these assumptions are corrected we find agreement. The two assumptions made by NV are:

\begin{enumerate}
    \item NV claim that the Hawking temperature in string units ($[T_H]_{\mathrm{string}}$) is  $\CJ_0$. We disagree. $\CJ_0$ is the local temperature experienced by generic chords confined to the stretched horizon. It is related to the Hawking temperature (seen at the pode)
    by the blue-shift factor $p$. The Hawking temperature in string units is, $$\boxed{[T_H]_{\mathrm{string}}= \frac{\CJ_0}{p}}.$$ This also follows from the analysis of Tomperature \cite{Lin:2022nss} after a change of units from cosmic to string.

    \item NV assume, without any justification that we understand, that the bulk deficit angle $\alpha$ and the parameter $v$ in \eqref{piv} are equal to one another. We know of no reason for such an equality. Our own analysis, based on standard relations between mass, deficit angle, and energy (as well as the standard relation between horizon area and entropy that has been assumed throughout this paper) lead to the relation $$\boxed{v \ \sim \ p\alpha}$$ for small $\alpha.$ This has an interesting implication. Our discussion has mostly focused on phenomena in the ranges of cosmic and string scales. States with appreciable deficit angle are in the third range, i.e., near the maximum mass scale $M_{\mathrm{max}}.$ What $v \sim p\alpha$ tells us is that these massive states with significant back-reaction reside in the non-gaussian tails of the DSSYK$_{\infty}$ energy spectrum.

\end{enumerate}
	
We should point out that we have addressed only a small fraction of the material in \cite{Narovlansky:2023lfz}. Most of that paper attempts to define and calculate correlation functions dual to bulk propagators at the pode. NV claim to reproduce the functional form of de Sitter correlators from SYK calculations. At the present time we have nothing to say about this other than that it is an important problem.

In this paper we have also analyzed the various concepts of temperature that appear in the DSSYK$_{\infty}$/dS duality. In particular, we have studied the temperature $T_{\mathrm{chord}}$ experienced by generic chords, which explains some properties of chords. $T_{\mathrm{chord}}$ is of order string scale implying that it is hot enough to have a large effect on chords. We believe that generic chords, which are confined to the stretched horizon, behave as if they are in a hot plasma, hot enough to ``melt" them into dissociated fermions. This explains why chord correlations factorize into products of single fermion correlation functions. Secondly it explains why the correlation functions exponentially decay. Fields propagating in a hot plasma can be over-damped and decay without oscillating which seems to be what chords do
		
		\section*{Acknowledgements}
		\quad We would like to thank H. Verlinde and V. Narovlansky for helpful discussions regarding their recent paper and H. Lin for helpful discussions about chord operators. A.R. and L.S. are supported in part by NSF Grant PHY-1720397 and by the Stanford Institute of Theoretical Physics.
		
		\appendix
		
		\section{More on Bulk Mass}
		\label{massApp}
		\quad Mass is a bulk concept which can be defined in a number of equivalent ways, such as via the Schwarzschild-de Sitter metric or via the theory of fluctuations (see e.g. \cite{Susskind:2021omt,Chandrasekaran:2022cip} and references therein). For localized masses in $(2 + 1)$-dimensions, it can also be defined in terms of the conical deficit angle associated to the conical defect imparted at the mass's location. In this appendix we will explain these various definitions of mass and their relationship.
		
		\subsection{Schwarzschild de Sitter Space}
		\quad The static patch of a stationary localized mass is described by the Schwarzschild de-Sitter (SdS) metric 
		\begin{equation}
			\boxed{\mathrm{d}s^2 = -f_M(r)\,\mathrm{d}t^2 + \frac{\mathrm{d}r^2}{f_M(r)} + r^2\,\mathrm{d}\phi^2}
			\label{SdS}
		\end{equation}
		with blueshift factor $f_M(r)$ given by 
		\begin{equation}
			\boxed{f_M(r) = 1 - 8GM - \frac{r^2}{\ell_{\mathrm{dS}}^2}}
			\label{fM3}
		\end{equation}
		The scaling of the mass $M$ defined by \eqref{fM3} is chosen so that the bulk energy $E_{\mathrm{bulk}}$ defined by the first law $\mathrm{d}E_{\mathrm{bulk}} = T_H\,\mathrm{d}S_{\mathrm{bulk}}$ goes (for small masses) like \cite{Teitelboim:2001skl}
		\begin{equation}
			\boxed{\delta E_{\mathrm{bulk}} = -\delta M}
		\end{equation}
		
		In the coordinates of \eqref{SdS}, the cosmological horizon is at radius
		\begin{equation}
			\boxed{r_c(M) \equiv \ell_{\mathrm{dS}}\sqrt{1-8GM}}
		\end{equation}
		and $r$ runs from $0$ at the center of the static patch (the ``pode") to $r_c$ at the horizon. $t$ and $\phi$ run over their usual ranges. In particular, $\phi$ is periodic with $\phi \sim \phi + 2\pi$. 
		
		Note that, for $M \neq 0$, the cosmological horizon shrinks inward to radius $r_c < \ell_{\mathrm{dS}}$, lowering the temperature as well as the entropy as we will now see.
		
		\subsubsection*{Temperature and Entropy of SdS}
		\quad The Gibbons-Hawking temperature $T_c$ of the cosmological horizon of \eqref{SdS} can be found in many ways. For example, it---or, rather, its inverse $\beta_c = T_c^{-1}$---can be found (the frame/units of \eqref{SdS}) as the periodicity of the time coordinate in the Euclidean continuation of the SdS metric after we demand that this geometry be smooth at the horizon. Regardless of methodology, and independent of dimension, we find\footnote{Here is the derivation using smoothness at the Euclidean horizon: We can study the near-horizon limit by defining $\epsilon \ll 1$ via 
			\begin{equation}
				r = r_c + \frac{f'(r_c)}{4}\,\epsilon^2
			\end{equation}
			in terms of which (the longitudinal part of) the SdS metric reads
			\begin{equation}
				\mathrm{d}s_{\parallel}^2 = -\frac{f'(r_c)^2}{4}\,\epsilon^2\mathrm{d}t^2 + \mathrm{d}\epsilon^2 + O(\epsilon^4)
			\end{equation}
			If we make the rescaling
			\begin{equation}
				t_{\mathrm{Rindler}} = -\frac{f'(r_c)}{2}\,t
			\end{equation}
			(since, always, $f'(r_c) < 0$) this simply becomes the 2D Euclidean Rindler metric, which is smooth at $\epsilon = 0$ provided $\mathrm{i}t_{\mathrm{Rindler}} \sim \mathrm{i}t_{\mathrm{Rindler}} + 2\pi$, i.e. provided $\mathrm{i}t \sim \mathrm{i}t + \beta_c$ with 
			\begin{equation}
				\beta_c = -\frac{4\pi}{f'(r_c)}
			\end{equation}
			where $\mathrm{i}t$ etc. is shorthand for the Euclidean time coordinate in the Euclidean continuation of the given geometry.
		}
		\begin{equation}
			\boxed{T_c = -\frac{f_M'(r_c)}{4\pi}}
		\end{equation}
		or, specializing to $(2 + 1)$-dimensions, 
			\begin{equation}
			\boxed{T_c =  \frac{1}{2\pi\ell_{\mathrm{dS}}}\inp{\frac{r_c}{\ell_{\mathrm{dS}}}} = \frac{\sqrt{1-8GM}}{2\pi\ell_{\mathrm{dS}}}}
			\label{Tc3}
		\end{equation}
		
		The entropy of the cosmological horizon is given in the usual way\footnote{We can also get \eqref{entropy} more rigorously by following the logic of \cite{Teitelboim:2001skl}, using the semiclassical limit of the bulk canonical ensemble defined therein.} by
		\begin{equation}
			\boxed{S_c = \frac{\mathrm{Area}_c}{4G} =  \frac{2\pi r_c}{4G}}
			\label{entropy}
		\end{equation}
		and the entropy deficit (relative to empty de Sitter space, the state of maximum entropy) is given by 
		\begin{equation}
			\boxed{\Delta S \equiv S - S_c = \inp{1-\frac{r_c(M)}{\ell_{\mathrm{dS}}}}S}
			\label{DSfinite}
		\end{equation}
		
		For small masses, we have
		\begin{equation}
			\boxed{\frac{T_c}{T_{GH}}  = \frac{r_c}{\ell_{\mathrm{dS}}} \ \approx \ 1 - 4GM + O(M^2)}
		\end{equation}
		and
		\begin{equation}
			\boxed{\Delta S \ \approx \ \frac{M}{T_{GH}} + O(M^2)}
			\label{deficit}
		\end{equation}
		Note that the linear relationship \eqref{deficit} only deviates from the full relationship \eqref{DSfinite} by at worst\footnote{More precisely, we have that for $0 < M \leq M_{\mathrm{max}}$
		\begin{equation}
				\frac{1}{2} \ < \ \frac{\Delta S(M)_{\text{linear approximation}}}{\Delta S(M)_{\text{full relationship}}} \ \leq \ 1
		\end{equation}} a factor of $\frac{1}{2}$, so we can say
		\begin{equation}
			\boxed{\Delta S \ \sim \ \frac{M}{T_H}}
			\label{SMT}
		\end{equation}
		following the notation outlined in the introduction.
		
		\subsection{Localized Masses are Conical Defects}
        \label{defects}
		\quad As alluded to previously, the $(2 + 1)$-dimensional SdS metric \eqref{SdS} actually represents a ``normal" static patch with a conical defect (the avatar of a localized mass in $D = 3$) located at the pode. To see this, define $\upalpha$ by 
		\begin{equation}
			\boxed{\inp{1-\frac{\upalpha(M)}{2\pi}} 
			= \sqrt{1-8GM}}
			\label{aMfull}
		\end{equation}
		and define new coordinates $(\tilde{t},\tilde{r},\tilde{\phi})$  by
		\begin{equation}
			\boxed{\tilde{\phi} = \inp{1-\frac{\upalpha(M)}{2\pi}}\phi}
		\end{equation}
		and
		\begin{equation}
			\boxed{\tilde{t} = \inp{1-\frac{\upalpha(M)}{2\pi}}t, \qquad \tilde{r} = \frac{1}{\inp{1-\frac{\upalpha(M)}{2\pi}}}\,r}
		\end{equation}
		Note that $\tilde{\phi}$ obeys $\tilde{\phi} \sim \tilde{\phi} + \inp{2\pi-\upalpha(M)}$ so that, in these new coordinates, there is an explicit angular deficit of $\upalpha(M)$. 
		
		In terms of these coordinates, the metric reads 
		\begin{equation}
			\boxed{\mathrm{d}s^2 = -f_0(\tilde{r})\,\mathrm{d}\tilde{t}^2 + \frac{\mathrm{d}\tilde{r}^2}{f_0(\tilde{r})} + \tilde{r}^2\mathrm{d}\tilde{\phi}^2}
		\end{equation}
		Which is an otherwise ordinary and empty static patch but with a conical defect of deficit angle $\upalpha(M)$ at the pode. There is no conical deficit at the horizon in either coordinate system since the usual periodicity $t_E \ \sim \ t_E + \beta_c$ then implies the periodicity $\tilde{t}_E \ \sim \tilde{t}_E + 2\pi\ell_{\mathrm{dS}}$, which is indeed the periodicity in Euclidean time required for a ``normal" static patch without a deficit at the horizon. Note that the fact that there is an angular deficit at the pode is a manifestation of the fact that the pode and horizon are not in thermal equilibrium---a conical deficit (localized mass) represents a \emph{non equilibrium} configuration.
		
		Note that, for any value of $\upalpha$, the entropy difference is given by
		\begin{equation}
			\boxed{\Delta S = \frac{\upalpha}{2\pi}\,S}
			\label{DSfiniteA}
		\end{equation}
		
		We can now see that, for multiple reasons, the maximum mass that one can place in $(2 + 1)$-dimensional de Sitter space is given by
		\begin{equation}
			\boxed{M = M_{\mathrm{max}} \equiv \frac{1}{8G}}
			\label{MMax}
		\end{equation}
		 When $M = M_{\mathrm{max}}$, the geometry breaks down (the horizon shrinks to the pode and the conical deficit angle goes to $2\pi$) and the entropy difference is maximal.
		
		For small mass/angle the relationship \eqref{aMfull} simplifies to 
		\begin{equation}
			\boxed{\upalpha(M) \approx 8\pi GM + O(G^2M^2)}
		\end{equation}
		Note that this linear relationship only deviates from the full relationship \eqref{aMfull} by at a factor of at worst\footnote{Again---and for the same reasons as before---we have that for $0 < M \leq M_{\mathrm{max}}$
			\begin{equation}
				\frac{1}{2} \ < \ \frac{\upalpha(M)_{\text{linear approximation}}}{\upalpha(M)_{\text{full relationship}}} \ \leq \ 1
		\end{equation}} $\frac{1}{2}$, so we can say
		\begin{equation}
			\boxed{\frac{\upalpha(M)}{2\pi} \ \sim \ \frac{M}{M_{\mathrm{max}}}}
			\label{aM}
		\end{equation}
		
		\subsection{Probabilities for Fluctuations}
		\quad The bulk mass can also be understood in terms of its probability to spontaneously nucleate at the pode, which is given by (see e.g. \cite{Susskind:2021omt,Chandrasekaran:2022cip})
		\begin{equation}
			\boxed{P \ \sim \ e^{-\Delta S(M)}}
		\end{equation}
		with $\Delta S(M)$ defined as in \eqref{DSfinite}. 
        This agrees, via \eqref{deficit} with the probability for the mass to have been emitted from the horizon as a Gibbon-Hawking quantum \cite{Gibbons:1977mu} 
		\begin{equation}
			\boxed{P \ \sim \ e^{-M/T_H}}
		\end{equation}
		
		The probabilities for more general (possibly non-geometric) fluctutions are again given by\footnote{This is simply the law of detailed balance.}
		(see e.g. \cite{Susskind:2021omt,Chandrasekaran:2022cip})
		\begin{equation}
			\boxed{P \ \sim \ e^{-\Delta S}}
		\end{equation}
		where the generalized entropy deficit
		\begin{equation}
			\boxed{\Delta S \equiv S_{\mathrm{dS}} - S'}
			\label{DeltaS}
		\end{equation}
		denotes the difference between the entropy of the empty static patch and the entropy of the new state. We can use this, as well as the relationship \eqref{SMT} to \emph{define} the mass of a generic (possibly nongeometric) state via 
		\begin{equation}
			\boxed{M \ \sim \  -T_H\log(P)}
		\end{equation}
		where we have used $\sim$ since this definition may differ from the one implied by \eqref{SdS} by an overall $O(1)$ multiplicative factor as well as an additive term which is subleading in the semiclassical limit.
		
\section{The Semiclassical Limit}
		\label{semiclassical}
			\quad In the context of de Sitter space (or anti de Sitter space for that matter) there are two different limits that are sometimes called the ``semiclassical limit".  The more common one---which we will call the ``weak" semiclassical limit---is one in which gravity behaves semiclassically at large scales while matter remains fully quantum. This is what is usually meant in the literature by ``semiclassical gravity" and is what is meant by the present authors. In the DSSYK literature (see e.g. \cite{Narovlansky:2023lfz,Lin:2023trc}) the term ``semiclassical limit" is sometimes used to refer to a different limit in which \emph{all} degrees of freedom have small fluctuations and behave semiclassically. We will refer to this alternate limit as the ``strong" semiclassical limit. Conflating these two notions of semiclassical limit (i.e. ``weak" and ``strong") can lead to confusion, since \emph{both} limits can be found in the SYK model. The strong semiclassical limit occurs in the limit $N \to \infty$ with $p$ fixed, i.e. in the limit $\lambda \to 0$. The weak semiclassical limit by contrast occurs in the limit $N \to \infty$, $p \to \infty$ with $\lambda = 2p^2/N$ held fixed. 
			
			\subsection{A Motivating Example}
   
			\quad The semiclassical limit plays some role in both NV's arguments as well as in our own, so we'll take some time to explain the two versions of the semiclassical limit and their relation to one another using an example involving an action encoding the various scales that show up in de Sitter quantum gravity.
			
			To be specific, we will consider a $(2 + 1)$-dimensional theory with gravity minimally coupled to a scalar field with a cubic self interaction. The stringy nature of the matter is accounted for by making the cubic coupling nonlocal on the string scale (see \eqref{nonlocality} below). We'll work in terms of dimensionless coordinates $X^{\mu}$ 
            and will write the line element as
			\begin{equation}
				\boxed{\mathrm{d}s^2 = \ell_{\mathrm{dS}}^2\,\mathsf{g}_{\mu\nu}\,\mathrm{d}X^{\mu}\mathrm{d}X^{\nu}}
			\end{equation} 
			and the scalar field as 
			\begin{equation}
				\boxed{\text{scalar field} = \frac{\upphi}{{\sqrt{\ell_{\mathrm{dS}}}}}}
			\end{equation}
			so that the ``metric coefficients" $\mathsf{g}_{\mu\nu}$ and ``field variable" $\upphi$ are similarly dimensionless. 
			This is of course simply what is meant by ``working in cosmic units", see Section  \ref{units} above.
			
\begin{figure}[H]
			\begin{center}
				\includegraphics[scale=.7]{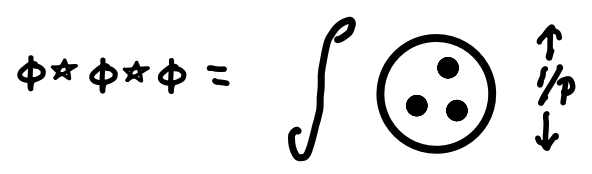}
				\caption{Schematic picture for the nonlocal coupling term $\upphi\star\upphi\star\upphi$, where the dots on the right hand side correspond to the positions of the three field insertions.}
				\label{nonlocal}
			\end{center}
		\end{figure}

			The gravity action is given by the Einstein-Hilbert action with positive cosmological constant $\Lambda_{(\mathrm{3D})} = \ell_{\mathrm{dS}}^{-1}$:
			\begin{equation}
				\boxed{I_{\mathrm{EH}}[\mathsf{g}] = \frac{\ell_{\mathrm{dS}}}{16\pi G}\int\mathrm{d}^3X\sqrt{|\mathsf{g}|}\inp{\mathsf{R} - 2}}
			\end{equation}
			Here $\mathsf{R}$ is the Ricci scalar of the dimensionless metric $\mathsf{g}_{\mu\nu}\,\mathrm{d}X^{\mu}\mathrm{d}X^{\nu}$. The action for the matter field is given by\footnote{This is a ``Yang-Mills" type frame for the interacting scalar field. We can relate this to the ``canonical" frame for the interacting scalar field---in which the field has a canonically normalized kinetic term---via the field redefinition $\upphi = g_s\upphi_0$, giving 
			\begin{equation}
					\boxed{I_{\mathrm{matter}}[\upphi] 
						= \frac{1}{2}\int\mathrm{d}^3X\sqrt{|\mathsf{g}|}\,\inp{\mathsf{g}^{\mu\nu}\partial_{\mu}\upphi_0\partial_{\nu}\upphi_0 + g_s^3 \,\upphi_0\star\upphi_0\star\upphi_0}}
			\end{equation}
			}
			\begin{equation}
				\boxed{I_{\mathrm{matter}}[\upphi] 
				= \frac{1}{2g_s^2}\int\mathrm{d}^3X\sqrt{|\mathsf{g}|}\,\inp{\mathsf{g}^{\mu\nu}\partial_{\mu}\upphi\partial_{\nu}\upphi +  \upphi\star\upphi\star\upphi}}
			\end{equation}
		with $g_s$ a coupling constant that we can consider to be the ``string coupling constant" due to the underlying stringy nature of the matter and its interactions. The symbol ``$\upphi\star\upphi\star\upphi$" is used to emphasize the nonlocality of the cubic coupling; specifically, we will take the coupling to be spread out over a coordinate distance 
		\begin{equation}
			\boxed{\Delta X \ \sim \ \frac{1}{p}}
			\label{nonlocality}
		\end{equation}
		(see fig. \ref{nonlocal}) with
		\begin{equation}
			\boxed{p \equiv g_s\sqrt{N}}
			\label{pBulk}
		\end{equation}
		Note that this ``bulk" definition of $p$ has simply been engineered by us to give answers reminiscent of the discussion in Section \ref{SepScales}.
            
		We can define the ``string length" $\ell_s$ as the dimensionful version of this nonlocality scale: 
		\begin{equation}
			\boxed{\ell_s \equiv \ell_{\mathrm{dS}}\cdot\Delta X = \frac{\ell_{\mathrm{dS}}}{p}}
		\end{equation}
		which mirrors/motivates the definition 
		\begin{equation}
			\boxed{\frac{\ell_s}{\ell_{\mathrm{dS}}} = \frac{1}{p}}
		\end{equation}
		given in \eqref{stringy} above. Defining as well
		\begin{equation}
			\boxed{N  \equiv \frac{\ell_{\mathrm{dS}}}{16\pi\hbar G} \ \sim \ S_{\mathrm{dS}}}
			\label{NBulk}
		\end{equation}
		we see that we can write the combined gravity-matter action as 
			{
			\begin{equation}
				\boxed{I[\mathsf{g},\upphi]
				= N\int\mathrm{d}^3X\sqrt{|\mathsf{g}|}\insb{\inp{\mathsf{R} - 2} + \frac{1}{2p^2}\inp{\mathsf{g}^{\mu\nu}\nabla_{\mu}\upphi\nabla_{\nu}\upphi + \upphi\star\upphi\star\upphi}}}
			\end{equation}
			\label{Npaction}
			}
			With these ``bulk" definitions \eqref{NBulk}, \eqref{pBulk} of $N$ and $p$, we find that 
			\begin{equation}
				\boxed{\lambda \equiv \frac{2p^2}{N} = g_s^2}
				\label{lambdags}
			\end{equation}
		      As an aside we note that, following Section \ref{SepScales}, we can also define the ``microscale" length $\ell_m$ by 
			\begin{equation}
				\boxed{\frac{\ell_m}{\ell_{\mathrm{dS}}} = \frac{1}{\sqrt{N}}}
			\end{equation}
			 This scale then satisfies 
			\begin{equation}
				\boxed{\frac{\ell_s}{\ell_m} = \frac{\sqrt{N}}{p} \ \sim \ \frac{1}{\sqrt{\lambda}}}
			\end{equation}
			
			\subsection{The Strong Semiclassical Limit}
			\quad Consider now the action \eqref{Npaction} in the limit $N \to \infty$ with $p$ held fixed. Since $p$ is a fixed, parameterically $O(1)$ number, the only control parameter is $N$ which multiplies the entire classical action 
			\begin{equation}
				\boxed{I = NI_{\mathrm{classical}}}
			\end{equation} 
			{
                \begin{equation}
				\boxed{I_{\mathrm{classical}}[\mathsf{g},\upphi]
                = \int\mathrm{d}^3X\sqrt{|\mathsf{g}|}\insb{\inp{\mathsf{R} - 2} + \frac{1}{2p^2}\inp{\mathsf{g}^{\mu\nu}\nabla_{\mu}\upphi\nabla_{\nu}\upphi + \upphi\star\upphi\star\upphi}}}
			\end{equation}
			}
			The bulk path integral is then dominated by the dominant stationary point---i.e. the ``classical solution" of--- $I_{\mathrm{classical}}$. It also follows from \eqref{lambdags} that in this same limit $g_s \to 0$: we see that when $N \to \infty$ at fixed $p$, all degrees of freedom become classical, i.e. cease to fluctuate. 
			
			The strong SCL has the following properties: 
			\begin{enumerate}
				\item The ratio $\ell_s/\ell_{\mathrm{dS}}$ is parameterically $O(1)$. The string scale and the cosmic scale do not separate. The nonlocalities implicit in the coupling $\upphi\star\upphi\star\upphi$ become parameterically cosmic in scale (though the nonlocality scale may still be \emph{numerically} small by taking $p$ large).
				
				\item Gravitational forces that vanish in the $N \to \infty$ limit are restored by finite $N$ corrections
				
								\item As explained in \cite{Susskind:2022bia}, this limit is precisely analogous to the 't Hooft limit of a gauge theory. The relation\footnote{In \eqref{aYM} above, $N_{\mathrm{YM}}$ and $g_{\mathrm{YM}}$ denote the number of colors and the Yang-Mills coupling of gauge theory respectively.}
				\begin{equation}
					\boxed{\alpha_{\text{'t Hooft}} \equiv N_{\mathrm{YM}}\,g^2_{\mathrm{YM}} = \mathrm{fixed}}
					\label{aYM}
				\end{equation}
				parallels the relation 
				\begin{equation}
					\boxed{p^2 = \lambda N = \mathrm{fixed}}
				\end{equation}
				Gauge theory diagrams which survive this limit are planar while SYK diagrams which survive this limit are melonic.
				
			\item 
                In fact an analog of the strongly-coupled 't Hooft limit (which is what is usually studied in the context of AdS/CFT) exists. It is defined by first taking the $N \to \infty$ limit with $p$ fixed and \emph{only then} letting $p \to \infty$. This corresponds to a $\lambda \to 0$ limit of DSSYK \cite{Lin:2023trc} which is called the ``large $p$ limit". It is still semiclassical in the strong sense but we additionally have that $\ell_s/\ell_{\mathrm{dS}} \to 0$.
			\end{enumerate}

                To re-emphasize, the \emph{strong} semiclassical limit represents a limit in which \emph{all} dimensionless couplings uniformly go to zero. These include matter couplings like the fine structure constant as well as the string coupling constant and the dimensionless gravitational coupling. In this limit the de Sitter entropy \eqref{Sbulk} goes to infinity \emph{and} all quantum fluctuations go to zero. In particular, the string scale in micro units diverges and string theory becomes not just perturbative but free. The string scale is then parameterically of order the (A)dS scale. In this limit second-quantized string theory becomes classical string field theory but first-quantized string theory (the worldsheet description) remains quantum, i.e. strings fluctuate.
			
			\subsection{The Weak Semiclassical Limit}
			\quad Now consider instead the true double-scaled limit 
			\begin{equation}
				N \to \infty \quad\mathrm{with}\quad \lambda = \frac{2p^2}{N} = \mathrm{fixed}
			\end{equation}
			The action \eqref{Npaction} naturally separates into two terms 
			\begin{equation}
				I[\mathsf{g},\upphi]
				= N\int\mathrm{d}^3X\inp{\mathsf{R}-2}
				+ \frac{1}{\lambda}\int\inp{\mathsf{g}^{\mu\nu}\partial_{\mu}\upphi\partial_{\nu}\upphi + \upphi\star\upphi\star\upphi}
			\end{equation}
			Gravitational fluctuations are supressed due to the factor of $N$ multiplying the Einstein-Hilbert action but matter fluctuations are not supressed, instead being controlled by the value of $\lambda$. In other words the matter theory is fully quantum. The string scale compared to the cosmic scale
			\begin{equation}
				\frac{\ell_s}{\ell_{\mathrm{dS}}} = \frac{1}{p}
			\end{equation} 
			goes to zero, indicating sub-cosmic locality. The string scale also remains finite on the microscale
			\begin{equation}
				\frac{\ell_{\mathrm{string}}}{\ell_m} = \frac{1}{\sqrt{\lambda}}
			\end{equation}
			This is analogous to saying that in $(3 + 1)$ dimensions the string scale stays finite in Planck units, which is of course the limit of sub-cosmic locality. 
			
		The finite $\lambda$ limit is much like the flat-space limit of AdS/CFT. The matter theory on scales $\lesssim \ell_{\mathrm{dS}}$ is identical to its flat-space limit. This is something that we predict to be true of DSSYK$_{\infty}$ as well \cite{Susskind:2022bia} but for the moment testing this conjecture remains difficult.

        The requirement for the weak semiclassical limit of gravity to be applicable while matter is described quantum-mechanically is just that the entropy is very large, i.e., that $N$ can be treated as almost infinite. Another way to say it is that there is a clear separation of scales.

        \subsection{Semiclassical Limits in DSSYK/dS}
        \quad As a matter of terminology, NV mean by ``the semiclassical limit" the limit in which $\lambda \to 0$, which is what we have called the ``strong" semiclassical limit here. By contrast, our own usage of the term ``semiclassical limit" refers to the limit in which $N \to \infty$ at fixed $\lambda$, which is what we have called the ``weak" semiclassical limit here. Both meanings of the SCL are legitimate but different. Although interesting, the issue of whether $\lambda \to 0$ defines a semiclassical limit is a side issue that has no connection with the disagreements that we are addressing.


\begin{thebibliography}{99} 
			
			\bibitem{Susskind:2021esx}
			L.~Susskind,
		``Entanglement and Chaos in De Sitter Space Holography: An SYK Example,''
			JHAP \textbf{1}, no.1, 1-22 (2021)
			doi:10.22128/jhap.2021.455.1005
			[arXiv:2109.14104 [hep-th]].
		
			
			\bibitem{Susskind:2022dfz}
			L.~Susskind,
			``Scrambling in Double-Scaled SYK and De Sitter Space,''
			[arXiv:2205.00315 [hep-th]].
			
			\bibitem{Susskind:2022bia}
			L.~Susskind,
			``De Sitter Space, Double-Scaled SYK, and the Separation of Scales in the Semiclassical Limit,''
			[arXiv:2209.09999 [hep-th]].
			
			\bibitem{Rahman:2022jsf}
			A.~A.~Rahman,
			``dS JT Gravity and Double-Scaled SYK,''
			[arXiv:2209.09997 [hep-th]].
			
			\bibitem{Verlinde}
			H. Verlinde,
			``A duality between SYK and (2+1)D de Sitter Gravity,"
			unpublished lecture,
			QGSC conference, Bariloche. 12/17/2019
			
\bibitem{Narovlansky:2023lfz}
V.~Narovlansky and H.~Verlinde,
``Double-scaled SYK and de Sitter Holography,''
[arXiv:2310.16994 [hep-th]].

        \bibitem{Susskind:2023hnj}
        L.~Susskind,
        JHAP \textbf{3}, no.1, 1-30 (2023)
        doi:10.22128/jhap.2023.661.1043
        [arXiv:2303.00792 [hep-th]].

\bibitem{Sybesma:2020fxg}
W.~Sybesma,
``Pure de Sitter space and the island moving back in time,''
Class. Quant. Grav. \textbf{38}, no.14, 145012 (2021)
doi:10.1088/1361-6382/abff9a
[arXiv:2008.07994 [hep-th]].

			
			\bibitem{Svesko:2022txo}
			A.~Svesko, E.~Verheijden, E.~P.~Verlinde and M.~R.~Visser,
			``Quasi-local energy and microcanonical entropy in two-dimensional nearly de Sitter gravity,''
			JHEP \textbf{08}, 075 (2022)
			doi:10.1007/JHEP08(2022)075
			[arXiv:2203.00700 [hep-th]].
			
			\bibitem{Susskind:2021omt}
			L.~Susskind,
			``De Sitter Holography: Fluctuations, Anomalous Symmetry, and Wormholes,''
			Universe \textbf{7}, no.12, 464 (2021)
			doi:10.3390/universe7120464
			[arXiv:2106.03964 [hep-th]].
			
			\bibitem{Bousso:1999xy}
			R.~Bousso,
			``A Covariant entropy conjecture,''
			JHEP \textbf{07}, 004 (1999)
			doi:10.1088/1126-6708/1999/07/004
			[arXiv:hep-th/9905177 [hep-th]].
			
			\bibitem{Susskind:2023hnj}
			L.~Susskind,
			``De Sitter Space has no Chords. Almost Everything is Confined.,''
			JHAP \textbf{3}, no.1, 1-30 (2023)
			doi:10.22128/jhap.2023.661.1043
			[arXiv:2303.00792 [hep-th]].
			
			\bibitem{Susskind:2023rxm}
			L.~Susskind,
			``A Paradox and its Resolution Illustrate Principles of de Sitter Holography,''
			[arXiv:2304.00589 [hep-th]].
			
			\bibitem{Lin:2022nss}
			H.~Lin and L.~Susskind,
		``Infinite Temperature's Not So Hot,''
			[arXiv:2206.01083 [hep-th]].

            \bibitem{Lin:2022rbf}
            H.~W.~Lin,
            JHEP \textbf{11}, 060 (2022)
            doi:10.1007/JHEP11(2022)060
            [arXiv:2208.07032 [hep-th]].
			
			\bibitem{Lin:2023trc}
			H.~W.~Lin and D.~Stanford,
			``A symmetry algebra in double-scaled SYK,''
			[arXiv:2307.15725 [hep-th]].
			
			\bibitem{Berkooz:2018jqr}
			M.~Berkooz, M.~Isachenkov, V.~Narovlansky and G.~Torrents,
			``Towards a full solution of the large N double-scaled SYK model,''
			JHEP \textbf{03}, 079 (2019)
			doi:10.1007/JHEP03(2019)079
			[arXiv:1811.02584 [hep-th]].
			
			\bibitem{Berkooz:2018qkz}
			M.~Berkooz, P.~Narayan and J.~Simon,
			``Chord diagrams, exact correlators in spin glasses and black hole bulk reconstruction,''
			JHEP \textbf{08}, 192 (2018)
			doi:10.1007/JHEP08(2018)192
			[arXiv:1806.04380 [hep-th]].

\bibitem{tHooft:1973alw}
G.~'t Hooft,
Nucl. Phys. B \textbf{72}, 461 (1974)
doi:10.1016/0550-3213(74)90154-0
			
   \bibitem{Maldacena:2016hyu}
			J.~Maldacena and D.~Stanford,
			``Remarks on the Sachdev-Ye-Kitaev model,''
			Phys. Rev. D \textbf{94}, no.10, 106002 (2016)
			doi:10.1103/PhysRevD.94.106002
			[arXiv:1604.07818 [hep-th]].

        \bibitem{Banks:2003cg}
        T.~Banks,
        [arXiv:astro-ph/0305037 [astro-ph]].

        \bibitem{Banks:2006rx}
        T.~Banks, B.~Fiol and A.~Morisse,
        JHEP \textbf{12}, 004 (2006)
        doi:10.1088/1126-6708/2006/12/004
        [arXiv:hep-th/0609062 [hep-th]].

        \bibitem{Fischler}
        W.~Fischler,
        ``Taking de Sitter seriously." Talk given at Role of Scaling Laws in Physics and Biology (Celebrating the 60th
Birthday of Geoffrey West), (2000)

        \bibitem{Dong:2018cuv}
        X.~Dong, E.~Silverstein and G.~Torroba,
        JHEP \textbf{07}, 050 (2018)
        doi:10.1007/JHEP07(2018)050
        [arXiv:1804.08623 [hep-th]].
			
	\bibitem{Gibbons:1977mu}
	G.~W.~Gibbons and S.~W.~Hawking,
	``Cosmological Event Horizons, Thermodynamics, and Particle Creation,''
	Phys. Rev. D \textbf{15}, 2738-2751 (1977)
	doi:10.1103/PhysRevD.15.2738
			
			\bibitem{Maldacena:2019cbz}
			J.~Maldacena, G.~J.~Turiaci and Z.~Yang,
			``Two dimensional Nearly de Sitter gravity,''
			JHEP \textbf{01}, 139 (2021)
			doi:10.1007/JHEP01(2021)139
			[arXiv:1904.01911 [hep-th]].
			
			\bibitem{Cotler:2019nbi}
			J.~Cotler, K.~Jensen and A.~Maloney,
			``Low-dimensional de Sitter quantum gravity,''
			JHEP \textbf{06}, 048 (2020)
			doi:10.1007/JHEP06(2020)048
			[arXiv:1905.03780 [hep-th]].
			
			\bibitem{Chandrasekaran:2022cip}
			V.~Chandrasekaran, R.~Longo, G.~Penington and E.~Witten,
			``An algebra of observables for de Sitter space,''
			JHEP \textbf{02}, 082 (2023)
			doi:10.1007/JHEP02(2023)082
			[arXiv:2206.10780 [hep-th]].
			
			\bibitem{Roberts:2018mnp}
			D.~A.~Roberts, D.~Stanford and A.~Streicher,
			``Operator growth in the SYK model,''
			JHEP \textbf{06}, 122 (2018)
			doi:10.1007/JHEP06(2018)122
			[arXiv:1802.02633 [hep-th]].
			
			\bibitem{Tarnopolsky:2018env}
			G.~Tarnopolsky,
			``Large $q$ expansion in the Sachdev-Ye-Kitaev model,''
			Phys. Rev. D \textbf{99}, no.2, 026010 (2019)
			doi:10.1103/PhysRevD.99.026010
			[arXiv:1801.06871 [hep-th]].
			
			\bibitem{Maldacena:2018lmt}
			J.~Maldacena and X.~L.~Qi,
			``Eternal traversable wormhole,''
			[arXiv:1804.00491 [hep-th]].
			
			\bibitem{Teitelboim:2001skl}
			C.~Teitelboim,
			``Gravitational thermodynamics of Schwarzschild-de Sitter space,''
			[arXiv:hep-th/0203258 [hep-th]].
			
			
\bibitem{Chandrasekaran:2022cip}
V.~Chandrasekaran, R.~Longo, G.~Penington and E.~Witten,
``An algebra of observables for de Sitter space,''
JHEP \textbf{02}, 082 (2023)
doi:10.1007/JHEP02(2023)082
[arXiv:2206.10780 [hep-th]].

			
		\end{thebibliography}
	\end{document}